\documentclass{elsart}
\usepackage{epsfig}
\usepackage{astrobib2}
\usepackage{graphicx}
\journal{New Astronomy}

\begin{document}
\def\bi#1{\hbox{\boldmath{$#1$}}}

\newcommand{\be}{\begin{equation}}
\newcommand{\ee}{\end{equation}}

\newcommand{\bea}{\begin{eqnarray}}
\newcommand{\eea}{\end{eqnarray}}

\def\lexp{\mathop{\langle}}
\def\rexp{\mathop{\rangle}}
\def\rexpc{\mathop{\rangle_c}}

\def\bi#1{\hbox{\boldmath{$#1$}}}
\def\bl{{\mathbf l}}
\def\nhat{\mathbf{\hat{n}}}
\def\epsl{E(\bl)}
\def\betal{B(\bl)}
\def\epslp{E(\bl')}
\def\betalp{B(\bl')}
\def\pee{P^{EE}_{l}}
\def\pbb{P^{BB}_{l}}
\def\peb{P^{EB}_{l}}
\def\pkk{P^{\kappa \kappa}_{l}}
\def\pek{P^{\kappa \kappa}_{l}}
\def\mapsize{1^{\circ} \times 1^{\circ}}
\def\sc{$\zeta$}

\begin{frontmatter}
\title{Mining Weak Lensing Surveys}

\author[Pton]{N.Padmanabhan},
\ead{npadmana@princeton.edu}
\author[Pton]{U.Seljak},
\ead{useljak@princeton.edu}
\author[CITA]{U.L.Pen}
\ead{pen@cita.utoronto.ca}

\address[Pton]{Department of Physics, Joseph Henry Laboratories, Princeton University, Princeton, NJ 08544, USA}
\address[CITA]{CITA, University of Toronto, 60 St. George Street, Toronto, Ontario M5S 3H8, Canada}

\begin{abstract}
We present a survey of the cosmological applications of the next
generation of weak lensing surveys, paying special attention to the
computational challenges presented by the number of galaxies,
$N_{gal}\sim$ 10$^{5}$. We focus on optimal methods with no pixelization 
and derive a multigrid $P^3M$ algorithm that performs the
relevant computations in $O(N_{gal} \log N_{gal})$ time. We test the
algorithm by studying three applications of weak lensing surveys - convergence
map
reconstruction, cluster detection and $E$ and $B$ 
power spectrum estimation using
realistic $\mapsize$ simulations derived from N-body simulations. 
The map reconstruction is able to reconstruct large scale features without 
artifacts.
Detecting clusters using only weak
lensing is difficult because of line of sight contamination and 
noise, with low completeness if one
desires low contamination of the sample. A power spectrum analysis 
of the convergence field is more promising and we are able to reconstruct 
the convergence spectrum with no loss of information down to the smallest
scales. The numerical methods used here can be applied to other data sets
with same $O(N\log N)$ scaling
and can be generalised to a sphere. 

\end{abstract}

\end{frontmatter}

\section{Introduction}

Mapping the matter distribution of the universe is one of the principal
aims of cosmology. The traditional approach to this problem has been the
use of galaxy surveys, the 2dF Galaxy Redshift Survey
\shortcite{2001MNRAS.328.1039C} and the Sloan Digital Sky Survey
\shortcite{2000AJ....120.1579Y} being the most relevant examples today.
However, galaxy surveys only map the luminous matter in the universe;
generalising to all forms of matter requires the additional assumption
that the luminous matter faithfully traces the total matter
distribution. On the other hand, weak lensing, or the coherent
distortion of the shapes of background galaxies by intervening matter,
requires no such assumption and is emerging as a powerful tool to map
the matter in the universe.

The possibility of lensing by large scale structure (LSS) was first
pointed out in pioneering work by \citeN{1967ApJ...150..737G} and has
since been theoretically and numerically studied by a number of authors
(\shortciteNP{1991MNRAS.251..600B}, \shortciteNP{1991ApJ...380....1M},
\shortciteNP{1992ApJ...388..272K}, \citeNP{1997A&A...322....1B},
\shortciteNP{1998ApJ...498...26K}, \shortciteNP{1997ApJ...484..560J},
\shortciteNP{2000Natur.405..143W}, \citeNP{2000ApJ...530..547J},
\shortciteNP{2000Natur.405..143W}). However, detecting this ``cosmic
shear'' had to wait for advances in imaging technology and has only
recently become possible. 
There now are a number of detections (\citeNP{2000MNRAS.318..625B},
\shortciteNP{2000A&A...358...30V},
\citeNP{2001ApJ...552L..85R}, 
\shortciteNP{2002ApJ...572...55H}) by various groups and with more
observations in progress or planned, this number will 
continue to grow. The next generation of lensing observations such as
the NOAO deep field, the CFHT legacy survey and the Deep Lens Survey 
will go beyond simply detecting cosmic shear but will map it over
large areas of the sky. 
Such large scale surveys will herald in an era of precision cosmology
for weak lensing. At the same time, these surveys bring with them the
same computational challenges that current CMB experiments and galaxy
surveys are facing. The number of background galaxies is
$N_{gal} \sim$ 10$^{5}$ to 10$^{6}$, making any brute force approach
prohibitively expensive. Analysing these surveys therefore requires the
development of algorithms that make the problem computationally
tractable. Furthermore, the algorithms must be able to handle aspects of
real data including noise, incomplete sampling, arbitrary cuts and so
on.

This work has two principal goals; the first is the development of an
algorithm that solves the computational challenges of large $N$
surveys, where $N$ can be number of galaxies or just pixelized intensity 
(as for the CMB). 
We then test this algorithm (described in detail in the
Appendix) on various applications of weak lensing surveys using
simulated data. The paper therefore serves the dual purpose of being a
test of our algorithm as well as a survey of the potential
of weak lensing surveys in cosmology.

We focus on three applications of weak lensing that we believe are the
most interesting for cosmology. The first of these, reconstructing the
matter distribution, is the basic goal of large lensing surveys. Lensing
measures the integrated line of sight matter distribution, or the
convergence ($\kappa$) map. Since this reconstruction includes all
matter, luminous and dark, such a map gives us a glimpse of the true
distribution of matter in the universe. Such maps could also be combined
with the high quality imaging data of the lensing survey to understand
the relationship between luminous and dark matter.
Weak lensing reconstructions have been considered under two broad
contexts, cluster mass reconstructions
(\shortciteNP{1993ApJ...404..441K}, \shortciteNP{1995ApJ...449..460K},
\shortciteNP{1995A&A...297..287S}) and LSS reconstructions
\shortcite{1998ApJ...506...64S}. Since the surveys we are considering in
this paper are large field surveys, we will focus on methods for the
latter, including effects of non uniform sampling, irregular boundaries
and noise.

The next scales of interest are the largest gravitationally bound
systems in the universe, clusters of galaxies. 
As clusters are the rarest and most massive of all structures, they
provide a sensitive probe of structure 
formation and the initial conditions of the universe. One such test is
the number density of clusters 
\shortcite{1998ApJ...504....1B} as a function of mass and redshift,
which 
depends sensitively both on the total matter in the universe as well as
the presence 
of dark energy and its properties. In order to use such tests, 
one needs to have a large catalogue of clusters, with well defined
selection criteria 
and completeness fractions. One approach is to use large photometric or
spectroscopic surveys of galaxies \shortcite{2001astro.ph.10307W} to
construct such catalogues; lensing provides us with an alternative
approach (\shortciteNP{2001AAS...199.1608H},
\shortciteNP{2001astro.ph.11490W}). An advantage to lensing is that it
detects all massive structures, whether they are luminous or not, while
other methods assume the presence of luminous matter. Since theory
predicts the total number of such objects, lensing allows for the most
direct comparison to predictions. Also, any 
possible existence of ``dark clusters'' (e.g.
\shortciteNP{2002A&A...388...68M}) would bias the latter methods, 
while lensing surveys would be unaffected. However, lensing has its own
disadvantages 
that must be understood, if it is to be reliably used.

This paper develops a method to detect clusters in lensing surveys, as
well as addressing the theoretical limitations of lensing for cluster
searches. It then characterises the completeness and reliability
(defined below) of such surveys. We also examine the potential of such
surveys to constrain the profile of the dark matter haloes associated
with such clusters. This work parallels and extends the work in
\shortciteN{2001astro.ph.11490W}, although we use different methods.

Finally, we address the measurement of the convergence power spectrum.
The power spectrum is one of a host of 
statistics that can be measured, but it has emerged as the statistic of
choice for cosmology, 
since many models predict a gaussian random field distribution on large
scales, for which the power spectrum 
contains all the information. Also, the power spectrum has a natural
interpretation from linear structure formation, and therefore is a
powerful probe of cosmology, especially in conjuction with other
astrophysical measurements (\shortciteNP{1997ApJ...484..560J},
\shortciteNP{1998ApJ...498...26K}, \shortciteNP{1999ApJ...514L..65H}).
The techniques for estimating the power spectrum are similar to those
already employed in CMB and galaxy redshift surveys. Here
we adapt them to weak lensing. 

The paper is organised as follows: We start by describing the features
of our algorithm. In $\S3$, we review the basic formalism of weak
lensing, while $\S4$ describes the simulations used in the paper. We
then examine each of our chosen weak lensing applications, image
reconstructions ($\S5$), cluster detections ($\S6$), and power spectrum
estimation ($\S7$ ). We conclude in $\S8$. The numerical algorithms used
in this paper are described in the Appendix.

\section{The Algorithm: Features}

As mentioned in the introduction, large weak lensing surveys have the
same computational challenges that the next generation of CMB and galaxy
survey analyses will face. Maximizing the weak lensing signal involves a
high number density of background galaxies distributed over large areas
of the sky, implying $N_{gal} \geq$ 100,000 galaxies. All the algorithms
described in this paper are written in terms of matrix
operations\footnote {Most modern cosmology algorithms are cast in this
form.}, and the computational challenge is in implementing these
$N_{gal} \times N_{gal}$ matrix operations efficiently. The goal of 
this paper is to show
that the challenge is tractable.

The actual numerical method is described in the Appendix; we discuss
the various features of the algorithm here.

\begin{enumerate}
\item {\it The algorithm works in real space}: There are a number of
reasons why a real space algorithm is preferable to a harmonic space
approach:
\begin{enumerate}
\item The noise properties of data are typically trivially representable
in real space.
\item Real data is not uniform, but involves a number of cuts that arise
from bad data, star removal, cosmic ray subtraction etc. In real space,
these cuts are trivially included. In harmonic space, these cuts would
introduce artifacts due to the non uniformity of coverage, and these are
generically difficult to correct for.
\item Since we do not work directly with Fourier modes 
there is no need to impose periodic boundary conditions.
Thus, there are no issues associated with aliasing of power 
from the modes on the scale of the survey. 
\end{enumerate}

\item {\it There is no pixelisation required}: A particularly simple
solution to the computational challenges of large data sets is to
pixelise the data if possible. There are a number of methods to do this,
and one can choose pixels that maximize the signal given a fixed number
of pixels. Pixelisation however does involve some degree of data loss at
all scales, and no information is preserved below the pixel scale. While
this may be acceptable for image reconstruction and power spectrum
measurements on large scales, 
it is undesirable for cluster searches, where small scale
information is essential. Furthermore, pixelisation is particularly problematic 
if the survey geometry is complicated.

\item {\it The time complexity is $N_{gal} \log N_{gal}$} : We have
tested our algorithm in the range $N_{gal} =$ 30,000 to 180,000 using a
single processor on a workstation. Within this range, the
computational time scales (effectively) linearly with the  number of
galaxies; the processing time ranged from a few minutes for the image
reconstructions and halo searches, to a few hours for the Fisher matrix
estimation. We also note that the algorithms in this paper,
especially the power spectrum estimation, are readily parallelized.

\item {\it The algorithm is generalisable} : The algorithm does not use
any properties that are unique to weak lensing; most of the methods
described here have either been borrowed from other applications, or
else are trivially generalisable. Consequently, this algorithm can be
used in other applications where large $N$ is a computational
restriction.

\end{enumerate}

\section{Lensing Formalism}
\label{lens_theory}

Let us begin by briefly reviewing the basic formalism of weak lensing to
establish notation and our conventions. The reader is referred to
\citeN{2001PhR...340..291B} for more details; our treatment closely
parallels the discussion in \citeN{2001ApJ...554...67H}.

The gravitational deflection of light can be described as a mapping
between a source (S) and image (I) plane. The mapping can be written as
\be
\delta x_{i}^{S} = A_{ij} \,\delta x_{j}^{I} \,\, ,
\label{eq:lensmapping1}
\ee
where $\mathbf{\delta x}$ is the displacement vector between two points
on a given plane. In the weak lensing regime, the mapping has the form
\be
A_{ij} = (1-\kappa) \delta_{ij} - \gamma_{1} \sigma_{3} - \gamma_{2}
\sigma_{1} \,\, ,
\label{eq:lensmapping2}
\ee
where $\sigma_{i}$ are the Pauli matrices, $\kappa$ is the convergence,
and $\gamma$ is the two component shear. Naively, one might expect that
this mapping depends on three independent parameters. However, the three
components of $A_{ij}$ are not independent; the relation between them
most easily expressed in Fourier space (we will assume a small enough
patch of sky in this paper to ignore curvature effects),
\be
\gamma_{1} (\bl) = \kappa(\bl) \, \cos(2\phi_{\bl}) 
\,\,\,\,\,\, ;\,\,\, \gamma_{2} (\bl) = \kappa(\bl) \,\sin(2\phi_{\bl})
\,\, ,
\label{eq:k2g}
\ee
where $\phi_{\bl}$ is the direction of the $\bl$ mode. Within the weak
lensing approximation, the expectation value of the ellipticity is
proportional to the shear. The proportionality constant depends on the
definition of the ellipticity; we adopt
\be
\langle e \rangle  = \gamma \,\, .
\label{eq:def_ellip}
\ee

We now compute the various lensing two point statistics. To do so, we
Fourier decompose the shear field into the so-called $E$ and $B$ modes,
\bea
\gamma_{1}(\nhat) = \int \, \frac{d^{2}l}{(2\pi)^{2}} \, [ E(\bl)
\cos(2\phi_{\bl}) - B(\bl) \sin(2\phi_{\bl}) ]\,  e^{i\bl . \nhat}
\nonumber \\
\gamma_{2}(\nhat) = \int \, \frac{d^{2}l}{(2\pi)^{2}} \, [ E(\bl)
\cos(2\phi_{\bl}) + B(\bl) \sin(2\phi_{\bl}) ]\,  e^{i\bl . \nhat}
\eea
The two point statistics of these quantities are specified by their power
spectra,
\bea
\langle \,\epsl \, \epslp \, \rangle = (2\pi)^{2} \delta(\bl - \bl') \,
\pee \nonumber \\
\langle \, \betal \, \betalp \, \rangle = (2\pi)^{2} \delta(\bl - \bl')
\, \pbb \nonumber \\
\langle \, \epsl \, \betalp \, \rangle = (2\pi)^{2} \delta(\bl - \bl')
\, \peb \,\, .
\eea
We note that weak lensing by density perturbations only produces $E$
modes, while shot noise and systematic effects can produce both. In
terms of power spectra, $\pee = \pkk\, , \, \peb = \pbb = 0$ for
weak lensing, while $\pee = \pbb \,,\, \peb = 0$ for shot noise.
Systematic effects can produce all three power spectra. For the rest of
this paper, we will ignore $\peb$ since it is parity violating; we refer
to $\pee = \pkk$ as the $E$ mode power, while $\pbb$ is the $B$ mode
power.

Using the above expectation values and standard trigonometric and Fourier
identities, we calculate the various covariance matrices,
\begin{enumerate}
\item 
\be
S^{\gamma \gamma}_{(ij) (ab)} = \langle \gamma_{ia} \gamma_{jb} \rangle
= \int \, \frac{l\, dl}{4\pi} \, \sum_{X=\epsilon\epsilon,\beta\beta} \,
P_{l}^{X} I^{X}_{ab} \,\, ,
\ee
with 
\bea
I^{\epsilon \epsilon}_{ab} = \left[ \begin{array}{cc} 
				J_{0} + c_{4}J_{4} & s_{4} J_{4} \\
				s_{4} J_{4} & J_{0} - c_{4} J_{4} 
				\end{array} \right] \nonumber \\
I^{\beta \beta}_{ab} = \left[ \begin{array}{cc} 
				J_{0} - c_{4}J_{4} & -s_{4} J_{4} \\
				-s_{4} J_{4}  & J_{0} + c_{4} J_{4} 
				\end{array} \right] \,\, .
\eea
\item 
\be
S^{\kappa \gamma}_{(ij) (a)} = \int \, \frac{l\,dl}{4\pi} \, \pek I_{a}
\,\, ,
\label{skg}
\ee
with
\be
I_{a} = ( \,-2 c_{2} J_{2} \, , \, -2 s_{2} J_{2} \,) \,\, .
\ee
\item
\be
S^{\kappa \kappa}_{ij} = \int \, \frac{l \, dl}{4\pi} \, \pkk J_{0} \,\,
.
\ee
\end{enumerate}
For convenience, we have introduced the shorthand notation $c_{n} =
\cos(n \phi),\, s_{n} = \sin(n\phi), \, J_{n} = J_{n} (l\theta)$. The
displacement vector between two points $(i,j)$ is described in polar
coordinates by $(\theta,\phi)$; the components of the shear are $(a,b)$.
We will often omit the superscripts for notational simplicity when
referring to these matrices if they are implicit in the context.

\section{Simulations}

\begin{table}
\begin{center}\begin{tabular}{cc}
\hline
Parameters & Value \\
\hline\hline
$\Omega_{M}$ & 0.3 \\
$\Omega_{\Lambda}$ & 0.7 \\
$\sigma_{8}$ & 0.9 \\
h & 0.7 \\
Boxsize & 239.5 Mpc/h \\
Particle Mass & $6.86 \times 10^{10}$ $M_{\odot}$/h \\
\hline
\end{tabular}\end{center}
\caption{The cosmological parameters used in the N-body simulations.}
\label{tab:simpar}
\end {table}

The simulated convergence maps used in this paper are derived from the
Virgo N-body simulations \shortcite{1998ApJ...499...20J}, run at the
Edinburgh Parallel Computing Center and the Computing Centre of the Max
Planck Society at Garching. These simulations use a parallel AP$^{3}$M
code
(\shortciteNP{1995ApJ...452..797C},\shortciteNP{1997NewA....2..411P}) to
evolve 256$^{3}$ particles from z=50 to z=0. The cosmological parameters
used in the simulation are in Table ~\ref{tab:simpar}. The dark matter
distribution is then projected onto a series of planes in the redshift
range z=2 to z=0. Dark matter haloes with a mass greater than 
 $10^{14}$ M$_\odot$
are identified with a FOF group finder with b = 0.2, forming the halo
catalogue we use in the rest of the paper.

\begin{figure*}
\begin{center}
\leavevmode
\includegraphics[width=5.0in, height=2.5in]{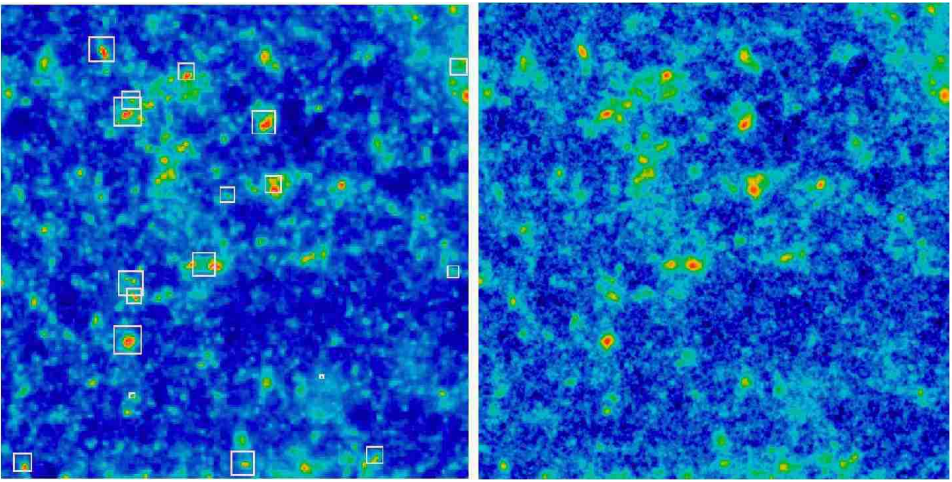}
\end{center}
\caption{Left: A simulated $\mapsize$ $\kappa$ map, assuming all the sources
are at z=1. The map was created using the method described in the text.
Haloes with masses greater than 10$^{14}$ M$_{\odot}$ are marked by squares with
the sizes scaled by $\kappa/\Sigma_{crit}$, where $\Sigma_{crit} =
\frac{c^{2}}{4\pi G} \frac{D_{s}}{D_{l}D_{ls}}$ and $D_s$, $D_l$, $D_{ls}$
are the
angular diameter  distances to the lens, source and between them, respectively.
The scale is between -0.06 and 0.26. Right: A WF reconstruction, assuming a
background density of n=25 galaxies/arcmin$^{2}$ and an unrealistically 
small intrinsic
scatter in the ellipticities of $\sigma_{int} = 0.4/100$.
}
\label{fig:kappa}
\end{figure*}

We compute the convergence map from these planes by using discrete
multiple plane lensing \cite{1992grle.book.....S},
\be
\kappa = \frac{3 H_{0}^{2}}{2} \Omega_{M} \sum_{i=1}^{N} g_{i} \,
\frac{\delta_{i}}{a_{i}} \, \Delta \chi_{i} \,\, ,
\label{eq:plane_lens}
\ee
where $\delta_{i}$ is the density perturbation at plane i and $g_{i}$ is
the geometrical distance ratio, given by
\be
g(\chi,\chi') = \frac{r(\chi') r(\chi - \chi')}{r(\chi)} \,\, ,
\ee
with $\chi$ the comoving distance and $r(\chi) = \sin_{K}(\chi)$ the
usual curvature distance. Two sets of convergence maps are created,
3.5$^{\circ} \times$ 3.5$^{\circ}$ maps assuming all the sources are at
z=1 and 2$^{\circ} \times$ 2$^{\circ}$ maps using the redshift
distribution,
\be
n(z) \propto z^{2} \exp(-z/z_{0}) \,\,\, , z_{0} = 0.4 \,\, ,
\label{eq:zdist}
\ee
to weight the planes in the sum. A representative $\mapsize$ subsection is
shown in Fig. \ref{fig:kappa}.
In order to create ``independent'' maps, the origin of the planes is
randomly chosen and we restrict ourselves to using $\mapsize$
subsections to create the galaxy catalogues. While the maps are
independent on small scales, this breaks down at larger scales, which
manifests itself in the high precision numerical experiments of Sec. 6. 

In order to generate galaxy catalogues, we create shear maps by Fourier
transforming the $\kappa$ maps and using Eq. \ref{eq:k2g}. In order to
eliminate artifacts from the periodic boundary conditions of the FFT,
only the central $\mapsize$ of the shear map is used. We randomly assign
galaxies positions, and bilinearly interpolate the shear map and use Eq.
\ref{eq:def_ellip} to compute their ellipticities. In addition, each
galaxy is given a random ellipticity drawn from a Gaussian distribution
with $\sigma = 0.4$. For the rest of the paper, we refer to the
intrinsic ellipticities of galaxies as ``noise''. In general, the noise
would also contain measurement errors, but we do not simulate these. Note
that our method simulates the effect of non-uniform coverage due to the clustering of background sources, although
we limit ourselves to Poisson clustering. Table \ref{tab:cattypes}
summarizes the different catalogues used in the various sections of the
paper.

\begin{table}
\begin{center}
\begin{tabular}{lcc}
\hline
Section & n & z distribution \\
\hline\hline
Image Reconstruction & 10,25,50 & yes \\
Clusters & 25,50,100 & yes \\
Power Spectrum & 25 & no \\
\hline
\end{tabular}
\end{center}
\caption{The table summarizes the types of galaxy catalogues used in
various sections of the paper. The background density of galaxies (per
arcminute$^{2}$) is given by n, while the third column indicates whether
a redshift distribution of the sources was considered or not. }
\label{tab:cattypes}
\end{table}

\section{Image Reconstruction}

One of the basic aims of a weak lensing survey is the reconstruction of
the 2D convergence field. A particularly simple nonparametric method for
this is Wiener Filtering (WF) \cite{1998ApJ...506...64S}. In the case
that the data are gaussian distributed, the WF estimator coincides with
the maximum posterior probability estimator
\shortcite{1995ApJ...449..446Z} and is therefore, optimal. Even when the
data are not gaussian, this estimator still minimizes the variance (as
defined below) among all linear estimators; however, it is no longer
guaranteed to be optimal. For LSS reconstructions, the deviations from
gaussianity on large scales are expected to be small, and so WF is, in
an appropriate sense, the best that one can do.

\subsection{Theory and Implementation}
To derive the estimator, let us organise the ellipticities of N galaxies
into a 2N - vector, $\mathbf{e}$. The estimator for the convergence
$\mathbf{\kappa}$, at M points can be written as $\mathbf{\hat{\kappa}
= \Phi e}$. We wish to minimize the variance, 
\be
\langle (\mathbf{\kappa - \Phi e})^{t}  (\mathbf{\kappa - \Phi e})
\rangle
\ee
with respect to $\mathbf{\Phi}$. This gives
\be
{\mathbf \Phi} = \langle \mathbf{\kappa e^{t}} \rangle \langle 
\mathbf{e e^{t}} \rangle^{-1} = S^{\kappa \gamma} (S^{\gamma \gamma} + N)^{-1} 
\ee
where $S^{ab}$ are the covariance matrices of Sec. 2, and $N$ is the 
noise covariance matrix. We note that $S^{\gamma \gamma} + N$ is a 2N
$\times$ 2N matrix while $S^{\kappa \gamma}$ is an M $\times$ 2N matrix.

In addition, it is essential to generate an error estimate for the
reconstruction. We do this by generating mock catalogues created by
randomizing the galaxies' ellipticities and reconstructing the $\kappa$
map for each of these catalogues. The variance over $N_{MC}$ (=200 in
this paper) maps is a measure of the error of the reconstructed
convergence field. It is important to note that in generating the mock
catalogues, the galaxy positions are not altered, thereby explicitly
taking into account the non-uniform (here Poisson) sampling of the data.

Using WF requires knowing the exact power spectrum (in
the covariance matrices) for it to be optimal. For this one can just use
the estimate of the power spectrum from the
data using the methods of Sec. 6 \cite{1998ApJ...506...64S}. However, 
empirical tests
demonstrate that the reconstruction is relatively  insensitive to the
power spectrum used if it is roughly approximates the true
one, even though the estimator is no longer strictly optimal.

Finally, on the implementation of this algorithm and others in the
paper; first, it is not necessary to estimate $\kappa$ at the positions
of the galaxies. Indeed, it is more useful to estimate it on a uniform
grid; all the figures in this paper are reconstructions on a 256$^{2}$ or 512$^{2}$ grid. Second, while the estimator is theoretically simple,
implementating the required matrix operations numerically for $10^{5}$
galaxies and greater, is more of a challenge. There are a number of
possible solutions to this; we refer the reader to the appendix for our
implementation. 

\subsection{Results}

\begin{figure*}
\begin{center}
\leavevmode
\includegraphics[width=5.0in,height=1.25in]{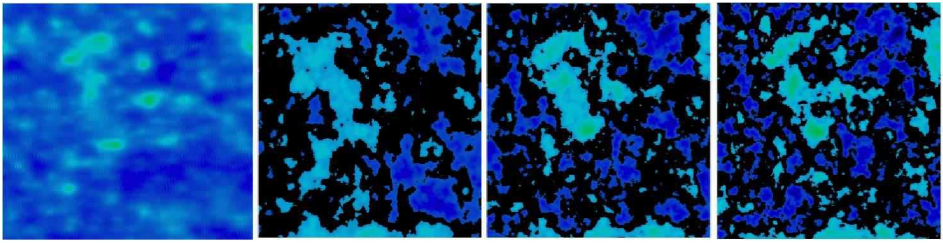}
\end{center}
\caption{ 
On the left a filtered 
version of the original map is shown, using a WF appropriate for 3rd panel. 
The right three panels show WF reconstructions of the $\kappa$ field of Fig.
\ref{fig:kappa}, 
where only those pixels where the measured signal exceeds the noise are shown. Bright (blue) 
are overdensities, dark (blue) underdensities, while black are pixels where signal is below noise.The
density of background galaxies is
10, 25, 50 galaxies/arcmin from left to right, respectively. The intrinsic scatter in the ellipticities
is $\sigma_{int} = 0.4$. The colour scale is identical to that in Fig. \ref{fig:kappa}.
}
\label{fig:noz_reconstruct}
\end{figure*}

\begin{figure}
\begin{center}
\leavevmode
\epsfxsize=3.0in \epsfbox{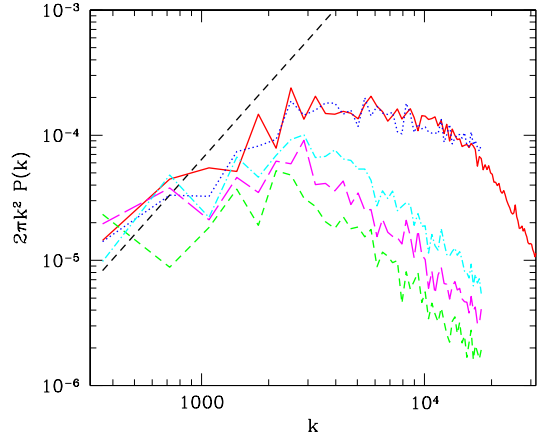}
\end{center}
\caption{ The power spectra for the WF reconstructions. The continuous
[red] line is the power spectrum of the original $\kappa$ map, while the
straight dashed [black] line is the noise power spectrum assuming n=25
galaxies/arcmin$^{2}$. The dotted [blue] line is the power spectrum of
the reconstruction in Fig.\ref{fig:kappa}, while the dashed lines are
the power spectra from Fig. \ref{fig:noz_reconstruct} (starting from
bottom to top, n=10,25,50 galaxies/arcmin$^{2}$).
}
\label{fig:sec1_power}
\end{figure}

The WF reconstructions of the field (Fig.\ref{fig:kappa}) for a variety
of background densities of galaxies and realistic noise are in Fig.
\ref{fig:noz_reconstruct}, while Fig. \ref{fig:kappa} shows the
reconstruction in the case where the noise has been reduced by a factor
of 100. The reconstructions resemble smoothed versions of the
original map. This can be understood by considering the WF operator in
Fourier space assuming diagonal noise. In Fourier space, the covariance
matrices are diagonal with $S^{\kappa \gamma} \sim S^{\gamma \gamma}
\sim \pkk$. The WF operator then simply weights each Fourier mode by
$\pkk /(\pkk + \hat{\sigma}^{2})$, where $\hat{\sigma}^{2}$ is the
amplitude of the noise power spectrum; i.e. every mode is simply
weighted by signal/(signal+noise). Considering Fig.\ref{fig:sec1_power},
we see that the principal effect of the WF is to act as a low pass
filter, removing modes greater than $l \sim$ 1000. Reducing the
level of the noise increases the cutoff frequency of this low pass
filter, as can be seen both from the images as well as the power
spectra.

The advantages of implementing the WF in real space are evident from the lack of artifacts at the edges of the reconstructions. The presence of an edge is irrelevant to a real space algorithm, and it correctly reconstructs structures at the edge of a field. This is not true for harmonic space approaches which are sensitive to the entire field, and generically produce artifacts at edges.

For completeness, we also considered the reconstructions of fields
assuming a redshift distribution of sources and verified that within our
implementation of the redshift distribution, there is no effect on the
reconstruction. This will continue to be true for more realistic
implementations of the redshift distribution except in the limit of
sparse sampling (low number of background galaxies). However, 
WF has only limited value in the sparse
sampling limit as is evident from the panels of
Fig.\ref{fig:noz_reconstruct}. 

\section{Clusters}

We turn to the construction and characterization of halo catalogues from
weak lensing data. Although weak lensing, being sensitive only to the
lensing mass, would appear to be the ideal method for constructing such
catalogues, it has its own theoretical limitations. We examine these
first, then turn to the construction of an optimal filter to detect
clusters, and test it. Finally, we address the issue of constraining
cluster profiles from weak lensing data.

\subsection{Theoretical Estimates}

\begin{figure}
\begin{center}
\leavevmode
\epsfxsize=3.0in \epsfbox{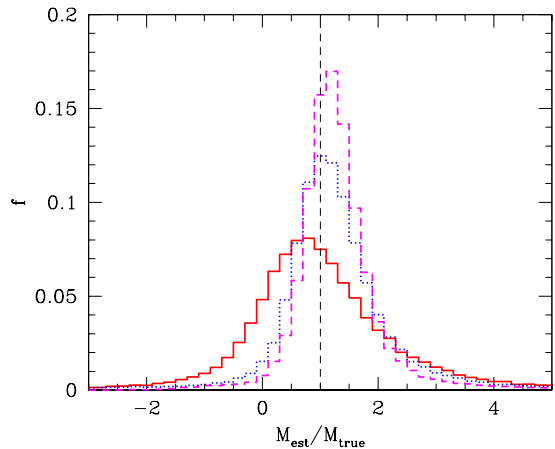}
\end{center}
\caption{ A histogram showing the fraction of haloes with ratio of $M_{lens}/M_{true}$from 200 realizations. The lensing mass was
computed by directly summing the $\kappa$ map upto 0.25 $\times$
[dashed/magenta], 0.5 $\times$ [dotted/blue], 1.0 $\times$ [solid/red]
the virial radius, $r_{vir}$, and multiplying with $\Sigma_{\rm crit}$
assuming cluster redshift is known. The halo finder mass is scaled
assuming an isothermal profile.
}
\label{fig:kappa_sum_hist}
\end{figure}

\begin{figure}
\begin{center}
\leavevmode
\epsfxsize=3.0in \epsfbox{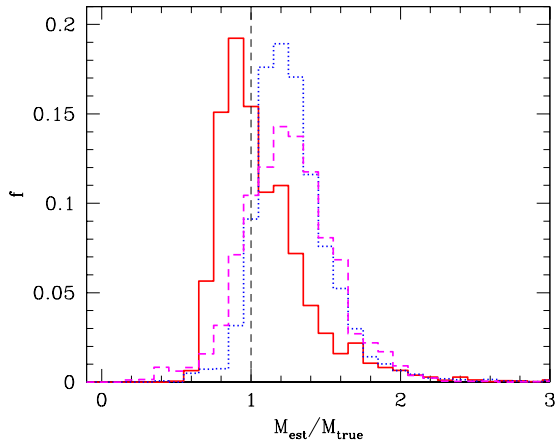}
\end{center}
\caption{ The same as Fig. \ref{fig:kappa_sum_hist} except only 50
realizations were considered and line of sight contamination was
removed.
}
\label{fig:zkappa_sum_hist}
\end{figure}

In order to understand the limitations of weak lensing halo searches, we
measure the expected signal from our catalogue of dark matter haloes,
with no noise added. We define the signal as the ratio
$M_{est}/M_{true}$, where $M_{est}$ is the lensing mass estimated from
the convergence map, while $M_{true}$ is the halo finder mass. The
lensing mass is estimated by integrating $\kappa$ outward from the halo
centre upto a fixed fraction of the virial radius\footnote{To get the
mass, one must multiply this by $\Sigma_{\rm crit}$. Equivalently, one
could scale the halo finder mass by this same ratio.}, while the halo
mass is determined by scaling the halo finder mass assuming an
isothermal density profile. The results are shown in Fig.
\ref{fig:kappa_sum_hist}. The slight trend in the mass ratio  
decreasing with radius is a result of clusters being steeper 
than isothermal in the outer parts of the cluster, such as in NFW profile. 

A feature of this plot is the presence of negative masses, i.e.
approximately 5\% of the haloes are undetectable. These negative masses
result from the fact that lensing measures the line of sight integrated
mass; low mass haloes can therefore be masked by underdensities. The
converse, low mass haloes being masked by heavier haloes, also occurs
and is responsible for the long tail of Fig. \ref{fig:kappa_sum_hist}.
Removing line of sight contamination (Fig. \ref{fig:zkappa_sum_hist}) by
considering each lensing plane individually (Eq. \ref{eq:plane_lens})
removes both the negative masses and the tail, verifying our
interpretation. This figure shows that the lensing masses
are systematically greater than the halo finder masses. This is the combination of two effects; the true profiles are not isothermal, and so using an isothermal profile to scale the halo finder masses will underestimate the mass. Also,
the lensing masses are 2D projected masses, while the halo finder
measures the 3D distribution.

Another related concern are filaments oriented along the line of sight
that could masquerade as haloes. In order to determine the degree of
contamination, we consider all pixels in our $\kappa$ maps that exceed a
particular threshold (we tried a few different thresholds, but our
results were insensitive to the particular choice). We then compute the
distance, $\Delta \theta$, between the pixel and the closest halo. This
is a measure of the correlation between the $\kappa$ overdensities and
the haloes. We find that all but a negligible fraction of overdensities
did not correspond ($\Delta \theta > 1$ arcmin) to real haloes, implying
that contamination due to filaments is insignificant.

\subsection{Cluster Detection: matched filters}

\begin{figure*}
\begin{center}
\leavevmode
\includegraphics[width=5.0in,height=6.5in]{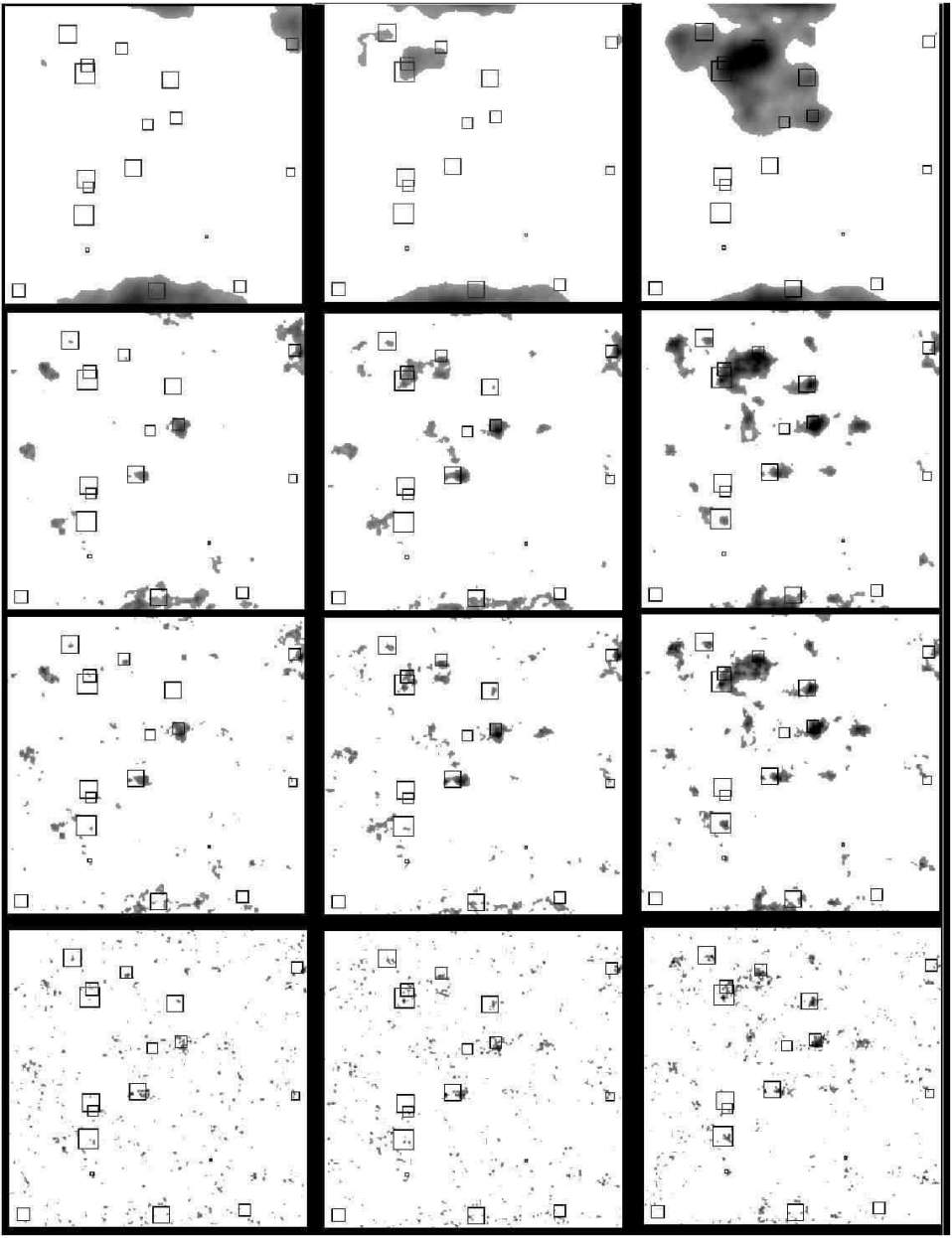}
\end{center}
\caption{ Cluster detections for a variety of scale lengths and galaxy
densities, using a threshold of $\sigma = 2$. The darkness is proportional to the significance of the detection, saturating for $\sigma > 5$. Haloes with masses over
$10^{14} M_{\odot}$ are marked by the boxes with the size of the box
dependent on the expected signal. The columns are different background
galaxy densities, n=25,50,100 galaxies/arcmin$^{2}$ from left to right.
The rows are different scale lengths, $\theta_{s}$ = 5.0,1.0,0.5 and 0.1
arcminutes from top to bottom. The realizations here did not include a
redshift distribution.
}
\label{fig:noz_clust}
\end{figure*}

\begin{figure*}
\begin{center}
\leavevmode
\includegraphics[width=5.0in,height=6.5in]{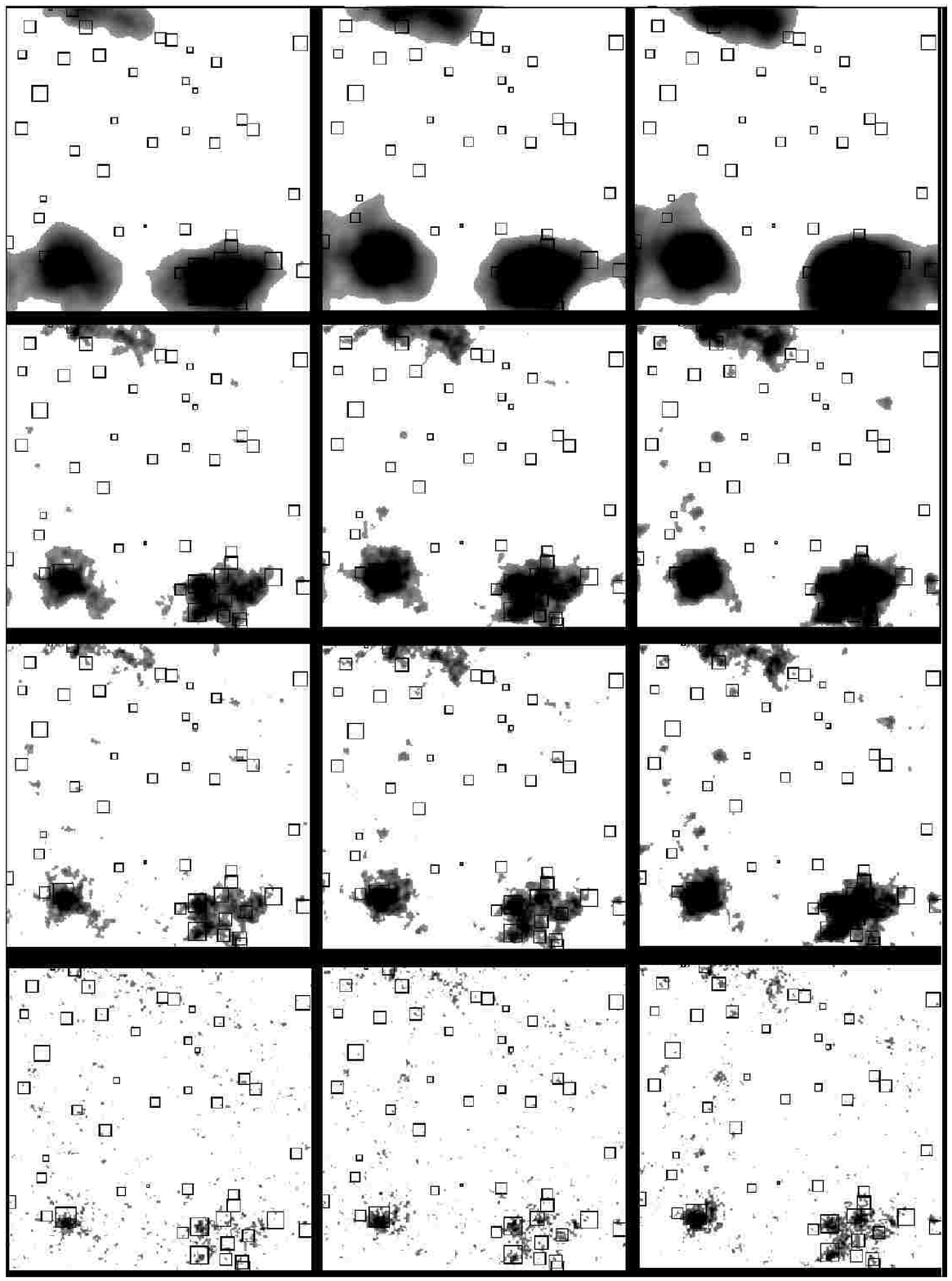}
\end{center}
\caption{ Same as Fig. \ref{fig:noz_clust} except that a redshift
distribution was used in the realizations.
}
\label{fig:z_clust}
\end{figure*}

The problem of halo detection can be formally stated as follows - given
an input signal with a spatial distribution $f({\mathbf x})$ and
amplitude $A$, the measured signal can be written as
\be
f'({\mathbf x}) = A \, f({\mathbf x}) + N({\mathbf x})
\ee
where $N({\mathbf x})$ is the noise. Assuming that the noise has zero
mean and and its statistical properties are spatially independent, the
minimum variance estimator, $\Phi$,  has the form (in Fourier space)
\cite{1996MNRAS.279..545H},
\be
\Phi({\mathbf k}) \propto \frac {f(\mathbf k)}{N({\mathbf k})} \,\, .
\ee
Specializing to the case of lensing, the estimator is given by
\be
\Phi \propto \frac{S'^{\kappa \gamma}}{N} \,\, .
\label{eq:matched_filt_matrix}
\ee
The normalization is usually determined by requiring that it be
unbiased. However, since we are interested in the ratio of the signal to
the noise, the normalization cancels and we leave it arbitrary. The
signal matrix, $S'^{\kappa \gamma}$, has the same form as the unprimed
matrix of equation \ref{skg}, 
except that the input power spectrum is replaced by the
Fourier transform of the halo profile. 

We now describe our halo detection algorithm. We start by convolving the
shear map with the matched filter; in practice, this involves
multiplying the data (organised into a vector) by the matrix in Eq.
\ref{eq:matched_filt_matrix}. Let us denote this convolved map by
$\tilde{\gamma} ({\mathbf x})$ - we must now determine whether the value
of $\tilde{\gamma}$ at a given point determines a cluster or not. In
order to do this, we start with the null hypothesis, i.e. there is no
halo at ${\mathbf x}$ and ask whether $\tilde{\gamma} ({\mathbf x})$ is
consistent with this. Under the null hypothesis, the expectation value
of $\tilde{\gamma} ({\mathbf x})$ is zero with variance $\sigma^{2}$. We
define a cluster detection if $\tilde{\gamma} ({\mathbf x}) > n \sigma$,
where $n$ is some chosen threshold. We compute $\sigma$ as in the
previous section, by randomising the galaxy ellipticities while 
keeping their positions fixed to create 200 
noise maps, and measuring
the variance of the resulting convolved maps.

We must emphasize a number of points at this stage. The first is that
for any given threshold, we expect to have points {\it not} associated
with a halo to have  $\tilde{\gamma} ({\mathbf x}) > n \sigma$; this
fraction of false detections will decrease with increasing $n$. If the
noise is Gaussian, then the fraction is known analytically; however, as
we will see in the following sections, the noise is not consistent with
being Gaussian. Indeed, the principal sources of noise come from
extraneous structures not associated with the halo; these structures do
not constitute a Gaussian random field. It is however possible to
calibrate the expected false fraction from simulations and include it in
theoretical analyses. We also re-emphasize that when computing the
variance, it is important to leave the galaxy positions unchanged and
only randomise the ellipticities. Only by leaving the galaxies'
positions unchanged will the signal, which is dependent on the
distribution of background galaxies, be properly estimated.

The cluster profile is still a free parameter. A particularly simple
choice is the singular isothermal sphere, which in projection is
$\propto \theta^{-1}$. This choice has the disadvantage of excessively
weighting the outer parts of the halo (the integrated profile is
logarithmically divergent), which are both noise dominated and
contaminated by external structures. Numerical experiments with it also
verify that it is suboptimal.

A simulation motivated choice is the NFW profile in projection
(\citeNP{1997ApJ...490..493N}, \citeNP{1996A&A...313..697B}). However,
the NFW profile has a $r^{-1}$ inner cusp, which is not resolved by our
simulations. Considering the NFW profile, convolved with the simulation
pixels, suggests a profile of the form \cite{2001astro.ph.10307W},
\be
f(\theta) = \frac{1}{(1+ \theta/\theta_{s})^{2}} \,\, ,
\label{eq:modif_nfw}
\ee  
where $\theta$ measures angular seperation on the sky. This profile has
the same aysmptotic $\theta^{-2}$ behaviour of the NFW profile; within
the scale radius $\theta_{s}$, it possesses a core. Note that we have
introduced an angular scale into the filter, and that there is
no way to theoretically choose this scale for all clusters, since they will
be at different distances. 
When analyzing real data one may of course choose a
projected NFW profile instead, which however still has a physical 
scale that cannot correspond to a single angular scale. 
We resolve
this problem by simply using a series of scale radii to perform the
reconstructions. 

The results for $\theta_{s} = $ 5.0, 1.0, 0.5 and 0.1 arcminutes are
shown in Figs. \ref{fig:noz_clust} and \ref{fig:z_clust}. As is evident
from the figures, the scale radius corresponds to a smoothing scale. A
larger scaling length results in fewer false detections, but tends to
coalesce seperate haloes into single structures and misses a greater
fraction of clusters. Decreasing the scale radius resolves seperate
structures better, but is also more susceptible to noise and extraneous
structures, as can be seen from the number of detected structures that
do not correspond to haloes.

\subsection{Completeness/Reliability}

\begin{figure}
\begin{center}
\leavevmode
\epsfxsize=3.0in \epsfbox{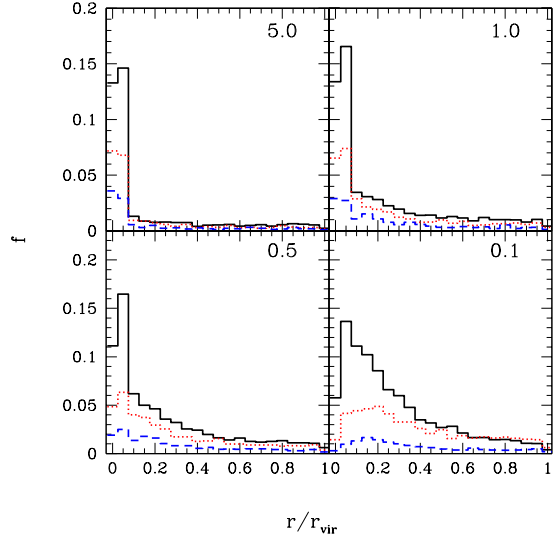}
\end{center}
\caption{ The completeness fraction for different thresholds and scale
lengths, computed from 200 realizations, with a background density of n
= 25 galaxies/arcmin$^{2}$. The panels are labelled by the scale length
in arcminutes. The solid [black], dotted [red] and dashed [blue] lines
represent thresholds of $\sigma = 2,3$ and $4$ respectively. The bin
size is 0.05 $r_{vir}$, implying the central completeness ($ r < 0.1
r_{vir}$) is contained in the first two bins. 
}
\label{fig:complete_n_25}
\end{figure}

\begin{figure}
\begin{center}
\leavevmode
\epsfxsize=3.0in \epsfbox{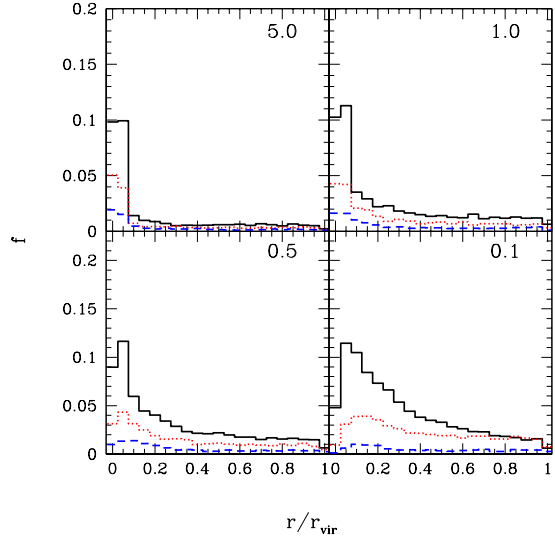}
\end{center}
\caption{ Same as Fig. \ref{fig:complete_n_25} except that a redshift
distribution was used in the realizations.
}
\label{fig:complete_z_25}
\end{figure}

It is useful to recall at this stage that we have two lists of data, a
list of haloes from the halo finder and a list of pixels whose halo
signal was above a certain threshold. There are two obvious questions
one can ask - what fraction of haloes have associated pixels, and what
fraction of pixels have associated haloes. We define the completeness as
the fraction of haloes that have their closest detected pixel a
specified distance away from them. Conversely, the reliability is
defined as the fraction of pixels that have their closest halo a
specified distance away. We note that the natural measure of distance
for completeness is in units of the virial radius, whereas the
reliability distance is measured as a physical angular distance. 

The completeness fractions, for three different thresholds, are shown in
Figs. \ref{fig:complete_n_25} and \ref{fig:complete_z_25}. We make the
following observations about these results:
\begin{enumerate}

\item As expected, increasing the threshold reduces the completeness
fraction. Note however that this happens at 2-3$\sigma$ level, 
so one cannot choose a high level of significance (thereby rejecting spurious 
detections with high confidence) 
and have a high level of completeness at the same time. 

\item The width of the completeness distribution is also seen to widen
as the scale length decreases. This can be understood heuristically by
noting that as the maps become more noise dominated, the point closest
to the halo may be unrelated to it, thereby broadening the distribution.
A crude modelling of the distribution in the limit of purely random,
independently distributed points obtains,
\be
f(x) \propto x\, (1-x^{2})^{n} \,\, ; \, x = r/r_{vir} \,\,,
\ee
which, for $n \sim 5$ resembles the distribution for the smallest scale
radius\footnote{The distribution assumed $n$ points distributed within
an area of $\pi r_{vir}^{2}$.}. Note however that this is meant to be
illustrative, and not a close approximation to the actual distribution.

\item A redshift distribution of the sources decreases the completeness fraction. This is the result of two competing effects - using a redshift distribution allows one to probe a larger volume of space, and one would hope to detect more clusters in that region. On the other hand, the redshift distribution results in a more non uniform signal which reduces the signal. 

\end{enumerate}

\begin{figure}
\begin{center}
\leavevmode
\epsfxsize=3.0in \epsfbox{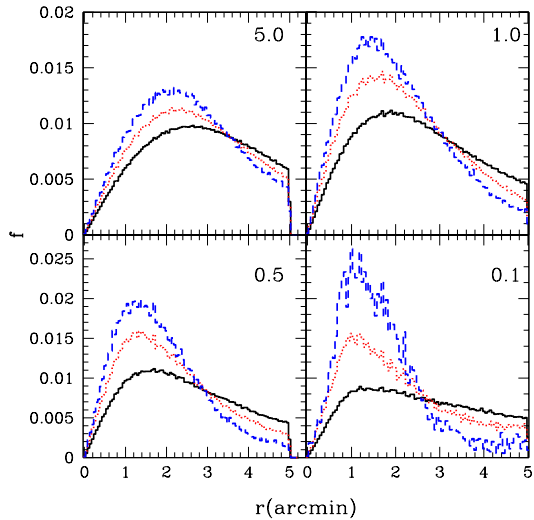}
\end{center}
\caption{ The reliability fraction for different thresholds and scale
lengths. The panels are labelled by the scale length in arcminutes. The
solid [black], dotted [red] and dashed [blue] lines represent thresholds
of $\sigma = 2,3$ and $4$ respectively.
}
\label{fig:false_n_25}
\end{figure}

\begin{figure}
\begin{center}
\leavevmode
\epsfxsize=3.0in \epsfbox{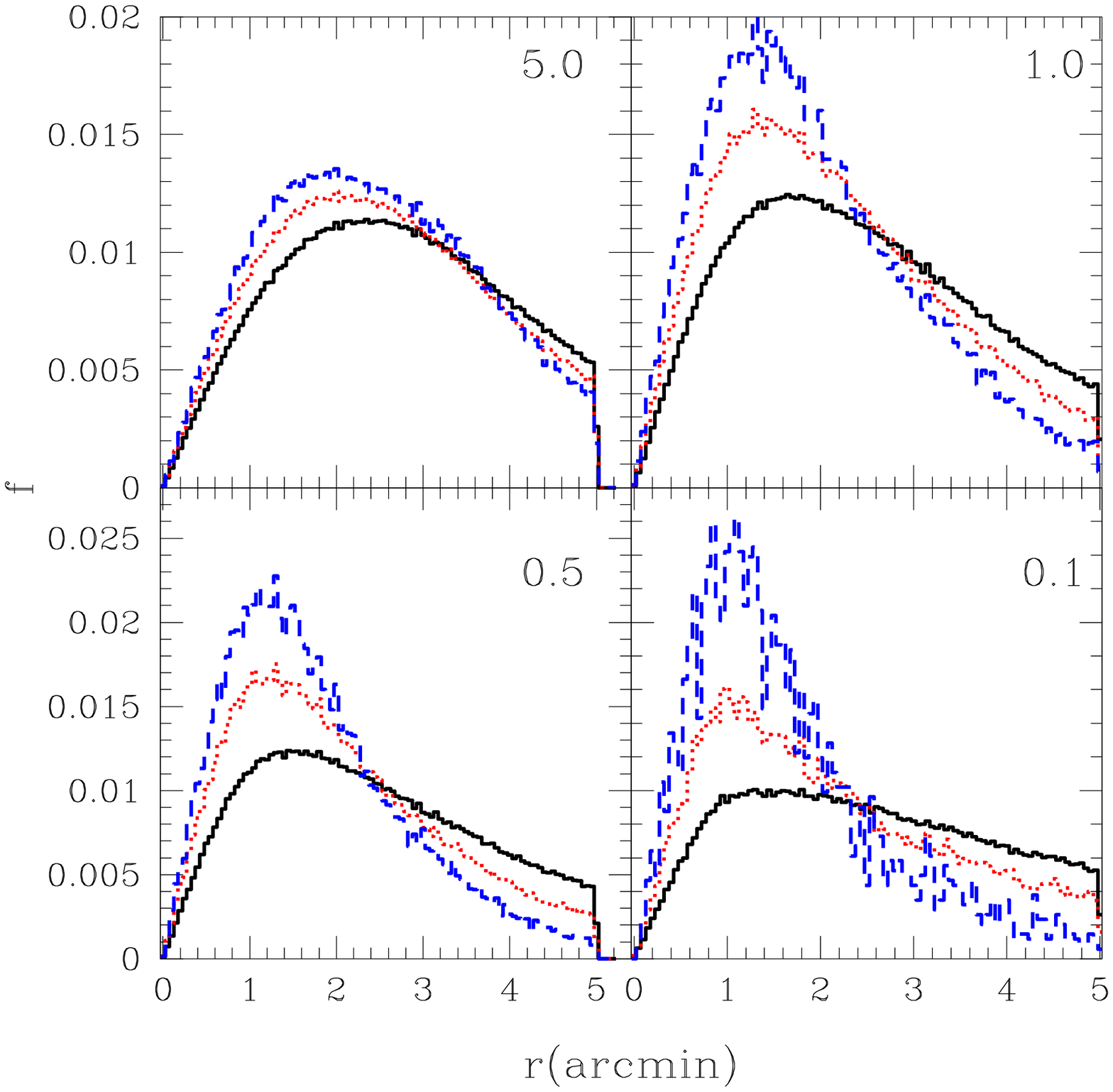}
\end{center}
\caption{ Same as Fig. \ref{fig:false_n_25} except that a redshift
distribution was used in the realizations.
}
\label{fig:false_z_25}
\end{figure}

\begin{table}
\begin{center}
\begin{tabular}{cccc}
\hline
$\theta$ (arcmin) & $\sigma$ =2 & 3 & 4 \\
\hline\hline
5.0 & 0.27 & 0.20 & 0.14 \\
1.0 & 0.23 & 0.12 & 0.06 \\
0.5 & 0.24 & 0.12 & 0.05 \\
0.1 & 0.35 & 0.22 & 0.07 \\
\hline
\end{tabular}
\end{center}
\caption{The fraction of detected pixels that did not have a
corresponding halo within 5 arcminutes, as a function of the scale
radius, $\theta_{s}$ and the threshold, $\sigma$. For simplicity, we
have only shown the figures for simulations without a redshift
distribution; the case with a redshift distribution is similar.}
\label{tab:false_fraction}
\end{table}

The corresponding reliability fractions are shown in Figs.
\ref{fig:false_n_25} and \ref{fig:false_z_25}, and the fraction of detected pixels that had
no corresponding halo within 5 arcminutes is in Table
\ref{tab:false_fraction}. Note that we have only shown the fractions for
the simulations without redshift distributions; including a redshift
distribution produces similar results. We note the following:
\begin{enumerate}

\item The reliability fraction becomes more peaked as
the threshold increases. 

\item The above trend becomes more pronounced as the scale radius
decreases, implying that a higher threshold is necessary to eliminate
noise at these scales. However, the number of detected pixels at both
high thresholds and small scales also steeply decreases, again pointing
to intermediate scales for optimal cluster detection.

\item Unlike the completeness, the reliability appears to be insensitive to
a redshift distribution.

\item Finally, the dropoff near the origin is a purely geometrical
effect, due to the fact that the differential area grows linearly with
radius.

\end{enumerate}

\begin{figure}
\begin{center}
\leavevmode
\epsfxsize=3.0in \epsfbox{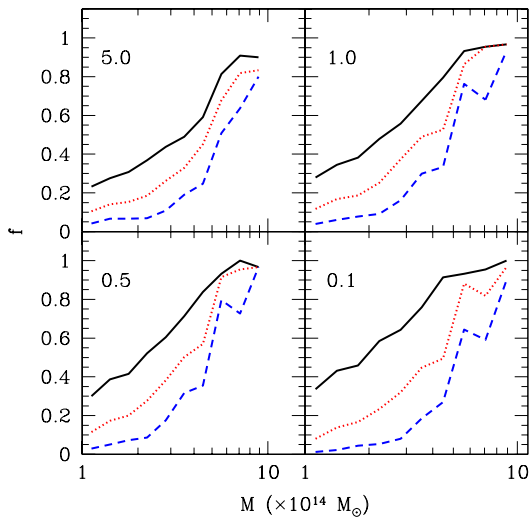}
\end{center}
\caption{ The completeness fraction (integrated to 0.2 r$_{vir}$) as a
function of halo mass. The panels are labelled by the scale length in
arcminutes. The solid [black], dotted [red] and dashed [blue] lines
represent thresholds of $\sigma = 2,3$ and $4$ respectively.
}
\label{fig:mass_complete_n_25}
\end{figure}

\begin{figure}
\begin{center}
\leavevmode
\epsfxsize=3.0in \epsfbox{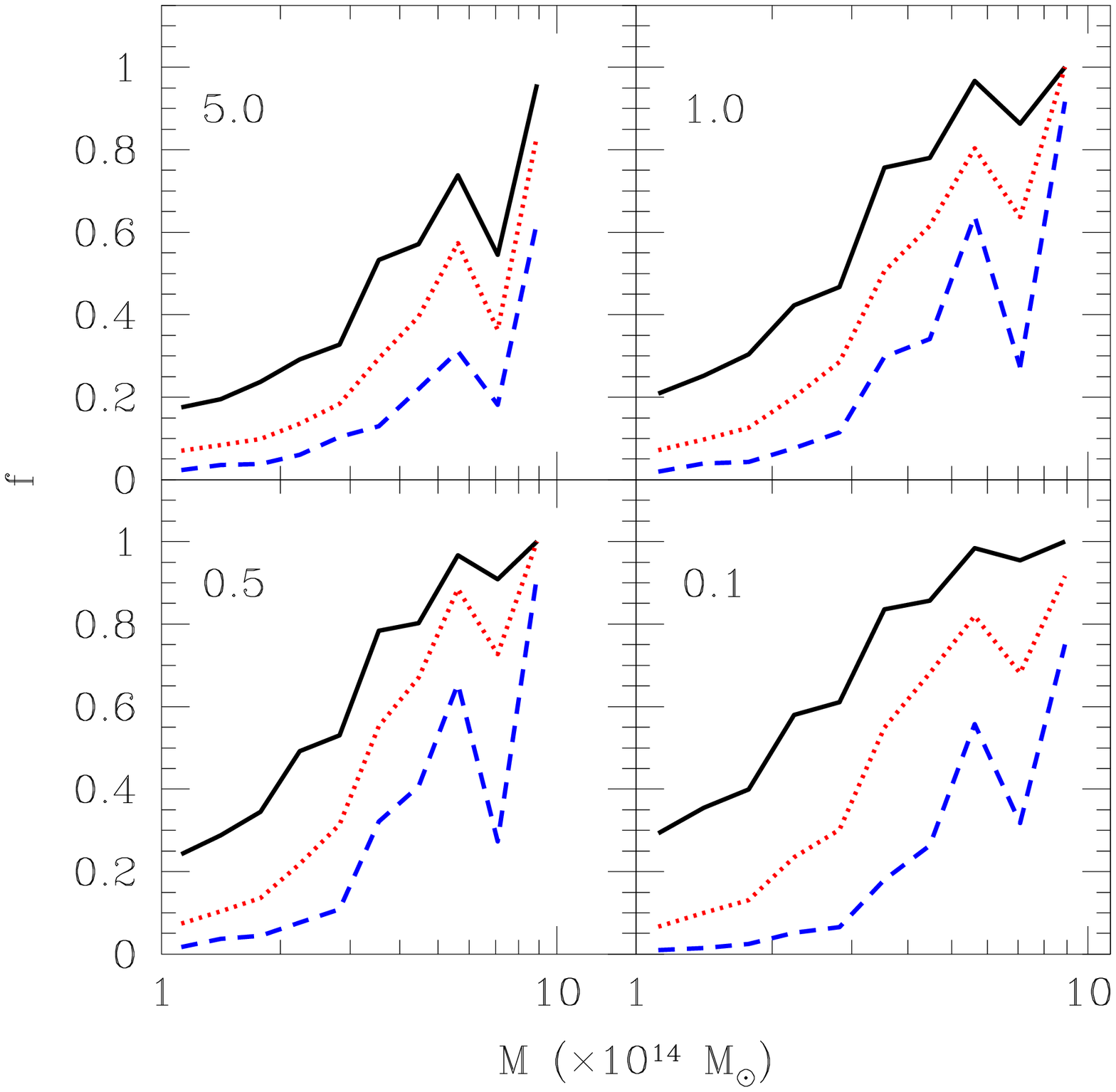}
\end{center}
\caption{ Same as Fig. \ref{fig:mass_complete_n_25} except that a
redshift distribution was used in the realizations.
}
\label{fig:mass_complete_z_25}
\end{figure}

It is also useful to consider completeness as a function of halo mass,
we do this in Figs. \ref{fig:mass_complete_n_25} and
\ref{fig:mass_complete_z_25}. The completeness plotted is an integrated
quantity - the fraction of haloes that have a detected pixel within 0.2
$r_{vir}$ of them. The completeness is a steep function of halo mass,
with the surveys being nearly 100\% complete at the high mass end, but
less than 5\% complete for the lowest masses. The noisy nature of these plots is simply an artifact of the fact that our simulations have few massive haloes.

An important observation to make is that the achieved completeness is
significantly lower than what one might have expected 
theoretically. This discrepancy
can be traced back to the assumption of a uniform signal in our
theoretical estimates. The Poisson clustering of the background galaxies
does not satisfy this assumption; indeed, configurations of background
galaxies make certain haloes undetectable. This would be even worse if
one included clustering of the background galaxies, and must be taken
into account when computing predicted cluster finding 
efficiency as a function of halo mass.

\subsection{Cluster Profiles}

\begin{figure}
\begin{center}
\leavevmode
\epsfxsize=3.0in \epsfbox{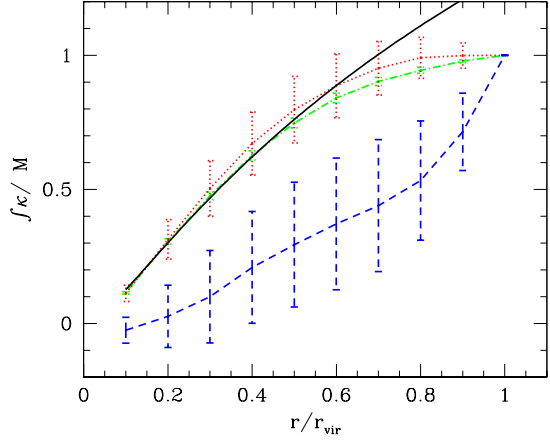}
\end{center}
\caption{ The integrated $\kappa$ profile created by stacking
clusters in three mass bins, 1-3 $\times$ [dashed/blue], 3-6 $\times$
[dotted/red], $>$ 6 $\times$ [dot-dashed/green] 10$^{14}$ M$_{\odot}$. The 
profile is arbitrarily normalized to unity at the virial radius. The
solid [black] line is arbitrarily normalized
c=4 projected NFW profile shown for comparison.
}
\label{fig:kappa_profile}
\end{figure}

\begin{figure}
\begin{center}
\leavevmode
\epsfxsize=3.0in \epsfbox{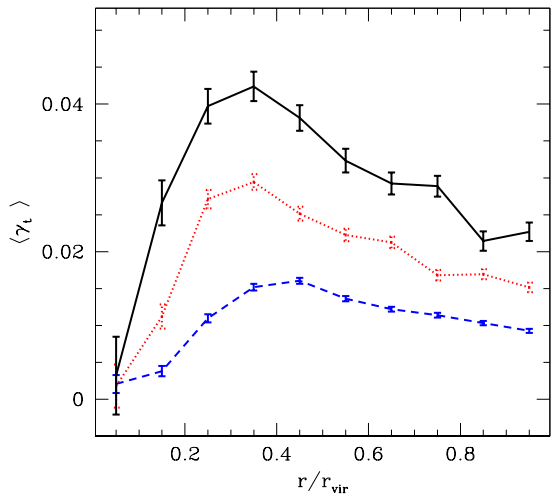}
\end{center}
\caption{Same as Fig. \ref{fig:kappa_profile}  except for the tangential
shear profile.
}
\label{fig:gamma_profile}
\end{figure}

\begin{figure}
\begin{center}
\leavevmode
\epsfxsize=3.0in \epsfbox{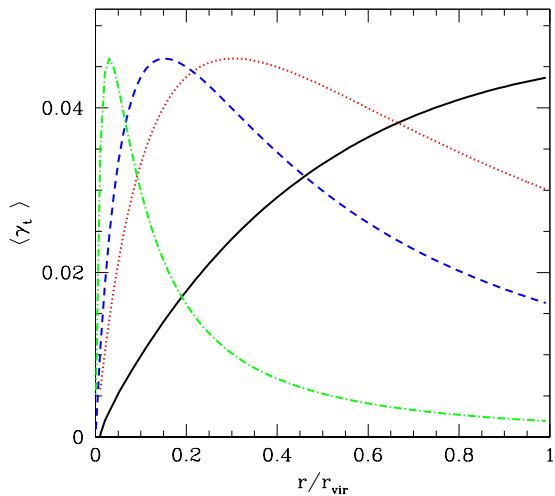}
\end{center}
\caption{The tangential profile for the profile in Eq.
\ref{eq:modif_nfw} with $\theta_{s}$ = 5.0 [solid/black], 1.0
[red/dotted], 0.5 [blue/dashed] and 0.1 arcminutes [dot-dashed/green].
The normalisation has been adjusted to agree with
Fig.\ref{fig:gamma_profile}, and the virial radius of the cluster has
angular size 2.5 arcminutes.
}
\label{fig:gamma_profile_exp}
\end{figure}

There has been considerable interest recently
\shortcite{2001ApJ...554..881S} 
in using weak lensing to statistically 
constrain cluster halo profiles. We examine this by considering both the
integrated mass profile (Fig. \ref{fig:kappa_profile}) and the
tangential shear profile (Fig. \ref{fig:gamma_profile}). The integrated
mass profile is simply given by summing our convergence map with no
noise added. For the shear profile, we use our simulated catalogues. The
ellipticity of each galaxy is decomposed, relative to the halo
center, into a tangential and radial component, which we then average in
radial bins to get the profile. Unlike the integrated mass profiles, the
shear profiles are estimated in the presence of noise.

We divide clusters into three mass bins, and average the profiles within
those bins assuming we know their central positions. 
This simulates the process of ``stacking'' clusters with centers known
from X-ray or optical data. 
The profiles are seen to be well constrained for the highest mass bin
and poorly constrained at low masses. 

We can also compare the measured shear profiles to what one would expect
if one assumed a convergence profile of the form in Eq. \ref{eq:modif_nfw}.
The expected profiles for the four scale radii we use are shown in
Fig.\ref{fig:gamma_profile_exp}. As we might have anticipated from the
results of the two previous subsections, the best agreement is when
$\theta_{s}$ is between 0.5 and 1.0 arcminutes. It is also important to
emphasise that the dropoff seen in Fig.\ref{fig:gamma_profile} is an
effect of the pixelisation of the map; a similar effect is seen in
Fig.\ref{fig:gamma_profile_exp} where the core radius mimics the effects
of pixelisation. If clusters had significant cores, then just
such an effect would be also physically expected. 

\section{Power Spectrum Estimation}

In this section we consider the measurement of the convergence power spectrum
from weak lensing data. The subject of the optimal measurement of the
power spectrum from noisy data has received a lot of attention, with
regards to the CMB and galaxy redshift surveys, and we will simply
import the techniques that have been developed and validated there to
weak lensing. It should be emphasized that weak lensing, unlike galaxy
redshift surveys, measures the {\it matter} power spectrum directly,
eliminating the complications of bias. 

Since the formalism for optimal weak lensing 
power spectrum measurement has been discussed
in detail elsewhere 
(see for eg. \shortciteNP{1998ApJ...506...64S},
\shortciteNP{1998ApJ...499..555T}), 
we will limit ourselves to a brief discussion (following 
the notation in \shortciteNP{2001ApJ...550...52P}) to establish the
formalism we will be using. A similar discussion that uses a different,
although related, approach is in \citeN{2001ApJ...554...67H}.

\subsection{Theory}

Let us parametrize the power spectrum by N$_{p}$ step functions such
that $P(l) = p_{i}$ for $l_{i-1} \leq l \leq l_{i}$, where $i$ ranges
from 1 to N$_{p}$. We can now arrange the $p_{i}$ into an
N$_{p}$-vector, ${\mathbf p}$. The problem now reduces to estimating
${\mathbf p}$ from the data ${\mathbf x}$, where ${\mathbf x}$ is a 2N
vector consisting of the galaxy ellipticities. Define the covariance
matrix of the data, $C = S^{\gamma \gamma} + N$ where $S^{\gamma
\gamma}$ is the signal covariance matrix (see Sec. 2) while $N$ is the
noise matrix. Recall that $N = N_{meas} + N_{int}$ where $N_{meas}$ is
the measurement noise, while $N_{int}$ is the intrinsic noise due to
galaxy ellipticities. This intrinsic noise is not known a priori and
must be estimated from the data, an issue we address at the end of this
section. 

Since the power spectrum is the sum of step functions, we rewrite the
covariance matrix,
\be
C = \sum_{i=1}^{N_{p}} p_{i} C_{i} + N = \sum_{i=0}^{N_{p}} C_{i} \,\, ,
\ee
where $C_{i} = \partial S^{\gamma \gamma} /\partial p_{i}$ for $1 \leq i
\leq N_{p}$, and we have defined $C_{0} = N$ and introduced a dummy
parameter, $p_{0} = 1$ for notational convenience. 

We now form the minimum variance quadratic estimators
(\shortciteNP{1997MNRAS.289..285H}, \shortciteNP{1997MNRAS.289..295H},
\shortciteNP{1997PhRvD..55.5895T}),
\be
q_{i} = \frac{1}{2} {\mathbf x}^{t} C^{-1} C_{i} C^{-1} {\mathbf x}
\ee
and group them into an N$_{p}$+1 vector, ${\mathbf q}$. The quadratic
estimators have the following properties,
\bea
\langle {\mathbf q} \rangle  = F {\mathbf p} \nonumber \\
\langle {\mathbf q \, q}^{t} \rangle - \langle {\mathbf q} \rangle
\langle {\mathbf q} \rangle^{t} = F 
\eea
where $F$ is the Fisher information matrix
(\shortciteNP{1997ApJ...480...22T} and refs. therein),
\be
F_{ij} = \frac{1}{2} tr \left[ C^{-1} C_{i} C^{-1} C_{j} \right] \,\,\,.
\label{eq:fisher}
\ee
The best estimate of the power spectrum can be written as 
\be
\mathbf{\hat{p}} = M ({\mathbf q} - {\mathbf b})
\label{eq:def_M}
\ee
where ${\mathbf b} = F_{i0}$, and we restrict all indices to run from 1
to N$_{p}$. The matrix $M$ is an arbitrary N$_{p} \times $ N$_{p}$ 
matrix with the property that the rows of $MF$ sum to unity and is
chosen such that our power spectrum estimates have certain desirable
properties. The choices for $M$ that we consider and their properties
are listed in Table \ref{tab:fisher_decorrelate}. Note that there is no
gain in information in this final step; the information content is the
same as what is in the quadratic estimators and the entire Fisher
matrix. However, if one only presents the diagonal of the Fisher matrix
(eg. errorbars in a plot), then it is important to ensure that the errorbars
are uncorrelated (the third choice). Our estimates $\mathbf{\hat{p}}$
now have the following properties,
\bea
\langle \mathbf{\hat{p}} \rangle = W {\mathbf p} = MF {\mathbf p}
\nonumber \\
\langle \mathbf{\hat{p} \hat{p}}^{t} \rangle - \langle {\mathbf p}
\rangle \langle {\mathbf p} \rangle^{t} = M F M^{t} \,\, ,
\eea
where we note that the rows of $W$ have the obvious interpretation as
window functions. 

\begin{table}
\begin{center}
\begin{tabular}{cc}
\hline
M & Properties \\
\hline\hline
$(\sum_{j=1}^{N_{p}} F_{ij})^{-1} \, \delta_{ij}$ & minimum variance, \\
& correlated errorbars \\
$F^{-1}$ & anti-correlated errorbars, \\
&  delta function windows \\
$[ \sum_{j=1}^{N_{p}} (F^{1/2})_{ij} ]^{-1} (F^{-1/2})_{ij}$ &
uncorrelated errorbars \\
\hline
\end{tabular}\end{center}
\caption{The various choices for $M$ (Eq. \ref{eq:def_M}) and their
properties.}
\label{tab:fisher_decorrelate}
\end{table}

\subsubsection{Choosing a prior}
In the preceding discussion, we have implicitly assumed a prior power
spectrum, and all the properties have rested on this prior being the
true power spectrum. This begs the question - how does one choose the
prior? Any reasonable guess to the power spectrum
works as an initial guess. The power spectrum estimated then can be used
as the prior, and the process can be iterated. 
Several authors
(\shortciteNP{1995PhDT........19B}, \shortciteNP{2001ApJ...550...52P} )
have explicitly verified that the choice of prior does not bias the
result, and therefore, in practice, only one or two iterations are
required.

\subsubsection{Measuring the intrinsic ellipticity scatter}
The intrinsic scatter in the ellipticities can be estimated by simply
computing the r.m.s. value of the ellipticity of the data. This will
produce the correct answer so long as the shear correlation length is
smaller than the size of the field considered. However, if the scatter
is underestimated, then the extra power will manifest itself as excess
power in the $E$ and $B$ modes. In order to correct for this, we follow
\citeN{2001ApJ...554...67H} and introduce a shot noise parameter,
$p_{noise}$, to the power spectrum, whose contribution to the power
spectrum is $C_{noise} = p_{noise} \delta_{ij}$. As with the power
spectrum, the excess power measured this way can be corrected in the
next iteration.

\subsection{Results}

\begin{figure}
\begin{center}
\leavevmode
\epsfxsize=3.0in \epsfbox{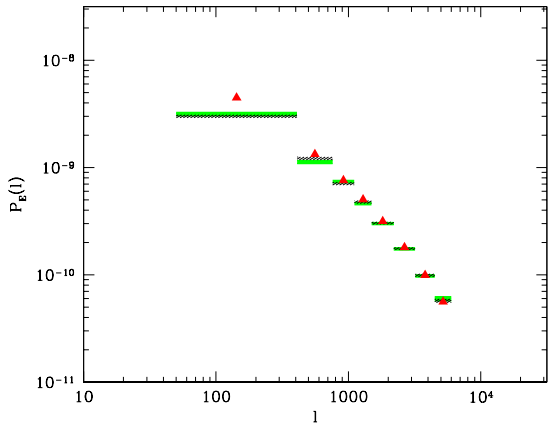}
\end{center}
\caption{ The $E$ mode power spectrum, averaged over 200 Gaussian
realizations. The [red] triangles show the input power spectrum, while
the heavily shaded [black] regions are the same power spectrum convolved
through the window functions. The width of the regions are the errors
predicted by the Fisher matrix. The lightly shaded [green] regions are
the estimated power spectrum, with the widths representing the errors
calculated from the realizations.
}
\label{fig:powspecge}
\end{figure}

\begin{figure}
\begin{center}
\leavevmode
\epsfxsize=3.0in \epsfbox{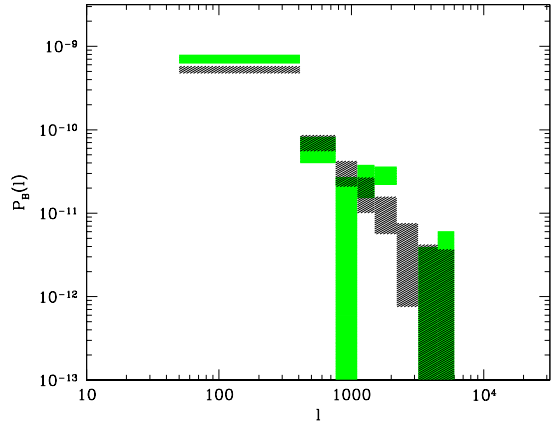}
\end{center}
\caption{ The same as Fig. \ref{fig:powspecge} except for the $B$ mode
power. Note that there was no input power; however, there is power
expected in the leftmost band due to leakage of power through the window
function.
}
\label{fig:powspecgb}
\end{figure}

The Fisher matrix formalism is only exact for Gaussian random fields,
and so, in order to test it, we simulate 200 gaussian random convergence
fields with a known $E$ mode power spectrum, and no $B$ mode power.
Figs. \ref{fig:powspecge} and \ref{fig:powspecgb} summarize the results.
The triangles are the input power spectrum, while the heavily shaded
boxes are the input power spectrum convolved with the window functions.
The width of the boxes are given by the Fisher matrix error estimates.
The lightly shaded boxes are the measured power spectrum, averaged over
200 realizations, with the width of the box given by the standard
deviation over the realizations. If the method is correct, then both the
power spectrum and the errors should agree. The agreement is best for
the $E$ mode power, where the input power was non-zero. We note that
although the input $B$ mode power was zero, we expect to see
power measured due to leakage from the $E$ modes through the window
functions. This is significant only for the leftmost band, that probes
scales larger than the survey\footnote{The modes were
included to prevent aliasing of power.}. We observe that the remaining
$B$ mode power is consistent with zero, as expected.

\begin{figure}
\begin{center}
\leavevmode
\epsfxsize=3.0in \epsfbox{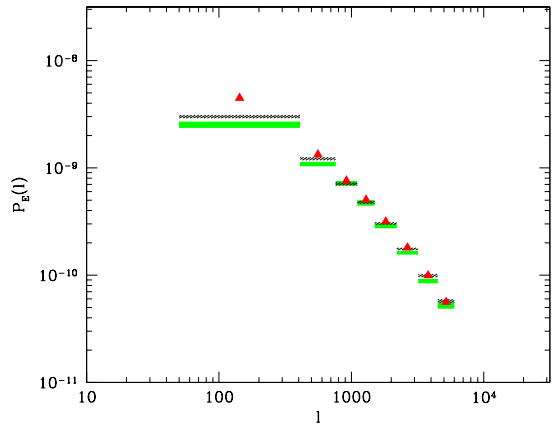}
\end{center}
\caption{ Same as Fig. \ref{fig:powspecge} except that the 200
realizations were drawn from N-body simulations instead.
}
\label{fig:powspecne}
\end{figure}

\begin{figure}
\begin{center}
\leavevmode
\epsfxsize=3.0in \epsfbox{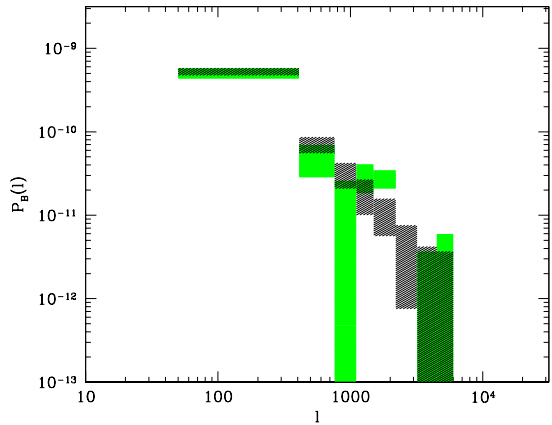}
\end{center}
\caption{ Same as Fig. \ref{fig:powspecgb} except that the 200
realizations were drawn from N-body simulations instead.
}
\label{fig:powspecnb}
\end{figure}

Figs. \ref{fig:powspecne} and \ref{fig:powspecnb} show the expected and
measured power for 200 realizations drawn from N-body simulations, which
allow us to test whether the breakdown of the gaussian approximation
biases our results. We find that
the Fisher matrix underestimates the true errors in the power
spectrum. A similar effect was found by \citeN{2001ApJ...554...67H}.
Although this is not a large effect, N-body simulations must be used to
calibrate the Fisher matrix errors when using the power spectrum
estimates to extract cosmological information. The second observation is
the discrepancy between the expected and measured $E$ mode power in the
leftmost bands. This is due to the fact that all our realizations were
derived from the same N-body simulations, and although we expect the
realizations to be independent on small scales, this will break down at
large scales. Fig. \ref{fig:powspecge} verifies this expectation. Note that these bands probe scales larger than those simulated; it is interesting that we correctly estimate the power to within a factor of two in these bins.

\begin{figure}
\begin{center}
\leavevmode
\epsfxsize=3.0in \epsfbox{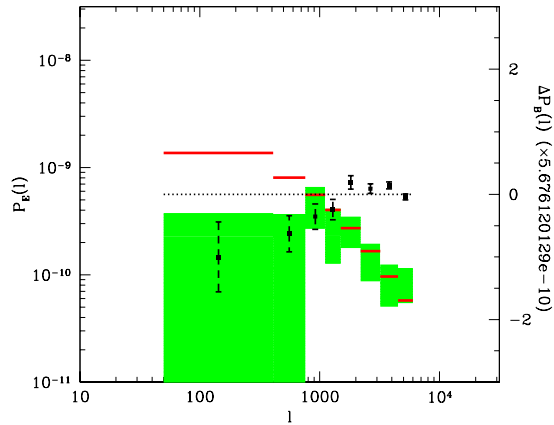}
\end{center}
\caption{ The measured power spectrum from a single N-body realization
using choice 1 of Table \ref{tab:fisher_decorrelate}. The solid lines
are the expected power, while the boxes are the expected power with
Fisher errors. The points are the deviation of the $B$ mode power from
zero, shown by the dashed line.
}
\label{fig:powspec_min}
\end{figure}

\begin{figure}
\begin{center}
\leavevmode
\epsfxsize=3.0in \epsfbox{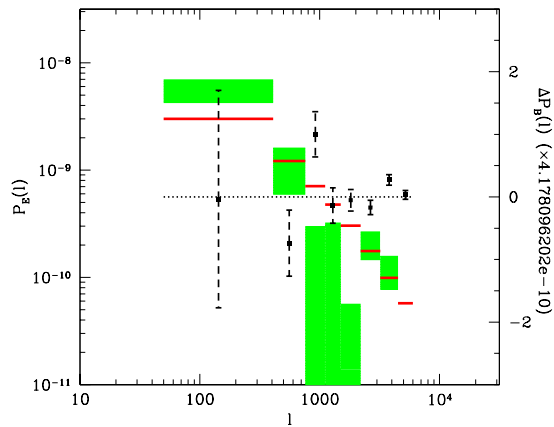}
\end{center}
\caption{ Same as Fig. \ref{fig:powspec_min} except that it uses choice
3 of Table \ref{tab:fisher_decorrelate}.
}
\label{fig:powspec_unc}
\end{figure}

All of the above results use the third choice of Table
\ref{tab:fisher_decorrelate} for $M$. Figs. \ref{fig:powspec_min} and
\ref{fig:powspec_unc} contrast the first and third choices for $M$. As
expected, the errorbars for the first are smaller, but are correlated,
as compared to the third. The effects of decorrelating are pronounced for the $B$ mode power; the points are randomly distributed above and below zero, while in Fig. \ref{fig:powspec_min}, they are clearly correlated. We therefore reemphasize the importance of decorrelating points when visually presenting data. We do not present the results for the second
choice since the errorbars are too large to be useful in this case. This
is a direct result of the fact that the window functions in this case
are delta functions.

\section{Discussion}

Weak lensing is emerging as an important tool in cosmology. One of its
principal advantages is that it probes the matter distribution directly,
making no assumptions about the dynamical state of the matter. This is
desirable both because it eliminates complications of interpretation,
but also because it gives us an opportunity to study the physical
processes underlying those assumptions.

In this paper, we considered three applications of large weak lensing
surveys. We summarise the results here, comparing it to previous work.

\subsection{Image Reconstruction}

The simplest goal of a weak lensing survey is to produce a map of the
distribution of matter in the universe. Wiener filtering
provides a simple, and almost
optimal reconstruction of the matter distribution, and is our method of
choice. 
An important characteristic of Wiener filtering is that it suppresses
power on scales with low signal to noise. For the weak lensing maps we
consider here, this scale is at $l \sim 1000$, while there is no power
at scales larger than $l \sim 360$, the scale of the survey. This
suggests the obvious design strategy of large area surveys to probe
large scales, with high background number densities $n \geq 25$ to
resolve features on small scales. However, Wiener
filtering has limited cosmological applicability because of its
inability to resolve smaller structures.

\subsection{Cluster Detection}

Clusters, as the most extreme structures in the universe, are a
sensitive probe of cosmology. Weak lensing has the advantage that it
searches for clusters directly as mass enhancements, independent of the
presence of luminous matter. However, in order to compare with
theoretical predictions, one must, in addition to compiling a catalogue
of clusters, understand the selection criteria and the completeness of
the catalogue. 

A disadvantage of weak lensing is that it measures the projected mass
distribution, and is therefore susceptible to contamination from
uncorrelated haloes in the line of sight. This is worst for the low mass
haloes, since they can be obscured by either more massive haloes or
underdensities. More massive haloes are less sensitive to this effect
and the theoretical maximum completeness approaches 100\%.

The philosophy that we adopt in this paper is that weak lensing cluster
searches will be done in conjunction with other measurements such as
galaxy, X-ray  or Sunyaev-Zel'dovich surveys. 
One could either imagine using weak lensing to identify candidate
clusters and verifying them with follow up measurements, or correlating
different measurements to remove false detections. With this in mind, we
develop an optimal filter approach to detecting clusters. Detections are
defined as those regions where the measured signal to noise ratio
exceeds a certain threshold. We then compute the completeness (the
number of haloes with a corresponding detection) and the reliability
(the number of detections with a corresponding halo). We observe that
completeness is strongly dependent on the mass of the cluster; for
cluster masses $>$ 6 $\times$ 10$^{14}$ M$_{\odot}$, our cluster samples
were virtually 100\% complete independent of the threshold used,
implying that one could construct a uncontaminated, yet complete sample,
by choosing a high enough threshold. At lower masses, the completeness
drops drastically as the threshold in increased. At low thresholds
however, the false detection rate rises to $\sim$ 25\%. This is a direct
result of the fact that our estimator is sensitive to spurious structure
masquerading as a halo at low thresholds. We should however note that
even for low thresholds, the completeness for low mass haloes is less
than 40\%.

The optimal filter defines a scale length that physically corresponds to
a smoothing scale. Excessive smoothing reduces the number of false
detections, since the noise is reduced but it also smooths away or
coalesces small sclae structures. As we reduce the smoothing scale, we
start to resolve smaller structures but are more susceptible to spurious
detections from extraneous structures that don't belong to any halo. The
optimal scale length is $\sim 1'$. However, some caution should be
exercised when interpreting this value since it may be affected by
pixelisation in the N-body maps.

This work parallels a similar study by \shortciteN{2001astro.ph.11490W},
although we use different methods to detect clusters.
\shortciteN{2001astro.ph.11490W} use both a simple Gaussian smoothing
with a smoothing scale of 1' - 2', and aperture mass measures
\shortcite{1996MNRAS.283..837S} with a scale of 1'-5' and a signal to
noise threshold varying from $S > 1$ to $S > 5$. As with our methods,
the completeness drops with increasing scale radius, with the maximum at
$\sim 1'$, although \shortciteN{2001astro.ph.11490W} do not consider
smaller scale lengths. In addition, they conclude that their catalogues
are complete for masses $> 5 \times 10^{14}$ M$_{\odot}$, consistent
with our results.

An extension of this is
the inclusion of a redshift distribution of the background sources.
Including this distribution does not qualitatively change any of the
trends one observes for, but reduces the overall completeness. The halo
catalogues, constructed assuming a redshift distribution, are complete
for masses $ > 8 \times 10^{14}$ M$_{\odot}$ . This is what one might
naively expect - the most massive haloes are unaffected by the redshift
distribution, while the signal from the less massive haloes is reduced
and a larger fraction of them are missed. The reliability is however
unaffected by a redshift distribution, because false detections are due
to noise and extraneous structure, and one would not expect those to be
affected by a redshift distribution.

There are two cautionary lessons that we draw from this work. The first
is that the optimal nature of matched filters is easily affected by
deviations from ideality. These are caused here by deviations from
circular symmetry due to extended haloes, as well as internal and external
substructure. The second is that the completeness of cluster surveys are
strongly affected by clustering of the background galaxies. Certain
haloes are rendered undetectable by the Poisson clustering of our
simulated background galaxies. Real background galaxies are not Poisson
distributed, but are correlated, 
and we expect this to have a worse effect. Therefore, all
analyses must take into account the clustering of source galaxies to get
a representative measure of the expected completeness of the survey.

We note that we have restricted ourselves here to consider the
information from lensing only. However, lensing surveys will produce high
resolution multi-colour images of the survey region. One might imagine
using both the lensing and imaging data simultaneously to detect
clusters; such an approach would naturally reduce the number of false
detections. Another advantage to using the imaging data jointly with the
lensing is that each complements the other - lensing would identify
haloes with no optical counterpart, while optical searches could detect
objects with low lensing signals.

\subsection{Power Spectrum}

The convergence power spectrum is a measure of the clustering of the
recent universe, a regime that was accessible until now only to redshift
surveys. In contrast with the galaxy power spectrum, the convergence
power spectrum is not affected by the complications of biasing, and
provides a direct measure of the matter power spectrum.

The formalism for estimating the power spectrum, borrowed from the CMB
and galaxy surveys, assumes that the convergence field is Gaussian
distributed. While this is not true for the true convergence field on
small scales, it is approximately true on the scales that lensing
probes. We have explicitly shown, using N-body simulations, that the
Gaussian approximation does not bias the estimators. However, the Fisher
matrix underestimates the error bars and although this is not a large
effect, it means that the error bars must be calibrated against N-body
simulations.
Our results are similar to those obtained by
\shortciteN{2001ApJ...554...67H} although our methods differ in details.
The main difference is that our method does not pixelise 
and so extracts all the information from the data. 

In concluding, we observe that the methods presented here are not
unique in their application to weak lensing. We have used methods 
developed for analysis of other data sets and adapted them to 
weak lensing. New numerical solutions presented here
may be adapted to other similar problems
in cosmology, particularly those where brute force evaluations are 
prohibitely expensive. In our application 
we have managed to reduce an instrinsic $O(N^3)$ 
numerical problem to $O(N\log N)$. The same methods can be used 
in other applications, such as the analysis of CMB temperature and 
polarization anisotropies and galaxy surveys. We expect that both
lensing and other areas of cosmology will benefit from the growing
synergy in the field.

We would like to thank SISSA, Trieste, where this work was begun, for its
hospitality.
We thank VIRGO collaboration for making N-body simulations 
available to the community. We would also like to thank Joe Hennawi, Catherine Heymans,
Christopher Hirata,  Kevin Huffenberger and David Spergel 
for useful discussions. N.P. was supported by a Centennial Graduate
Fellowship from Princeton University.
U.S. is supported by NASA, NSF CAREER, David and Lucille
Packard Foundation and Alfred P. Sloan Foundation. 
U.P. is supported by NSERC grant 72013704.
Computing infrastructure was provided in part by the Canada Foundation 
for Innnovation PScinet alphacluster.

\bibliography{paper}

\appendix
\section{Numerical Implementations}
An important feature of all the algorithms presented in this paper is
that they are explicitly written as linear algebra operations. The basic
building block of any implementation is therefore a routine to perform
matrix-vector multiplication. Unfortunately, the dimensionality of the
vector space, $N$, is given by the number of data points. A naive
application of the algorithms yields $N \sim N_{gal} \approx 100,000$
for which straight matrix-vector multiplication, an $N^{2}$ process,
becomes computationally impractical.

One can approach this problem in two ways, either by reducing $N$, or by
using properties of the matrices that appear to speed up the vector
multiplications. The former approach is equivalent to pixelising the
data and a number of pixelisation schemes have been suggested, ranging
from direct binning to ``optimal'' Karhunen-Lo\'{e}ve pixelisations of
the data.

We propose and implement an approach that is based on the latter
approach. We start by observing that all the matrix operations that we
require are of the form $C{\mathbf x}$ or $C^{-1} {\mathbf x}$ where $C$
is a correlation matrix of $\S3$. The latter operations can be recast as
direct matrix-vector multiplications by performing the matrix inversion
iteratively (we specify the exact algorithm below). We then observe that
the multiplication $C{\mathbf x}$ is simply a convolution of the data by
the appropriate correlation function, \sc,
\be
(C{\mathbf x})_{i} = \sum_{j} C_{ij} \, x_{j} = \sum_{j} \zeta({\mathbf
r}_{i} - {\mathbf r}_{j}) x_{j} \,\, ,
\ee
where the last identity follows from the fact that the correlation
function only depends on the seperation between the two points. The
problem is now explicitly translationally invariant and one can readily
apply the Fourier convolution theorem to perform the vector
multiplications efficiently. The asymptotic scaling is now $O(N \log N)$
instead of $O(N^{2})$, making $N \sim 10^{5}$ tractable on a workstation. 

The careful reader will no doubt point out that the use of Fourier
methods requires the data to be uniformly sampled, which our data is
not. We solve this problem by resampling the data onto a grid of
$N_{grid} \sim 4N_{gal}$ points where the additional factor of 4 comes
from the need to zero pad in 2 dimensions. The exact scaling of the
algorithm is therefore $N_{grid} \log N_{grid}$. We emphasise that
although an auxiliary grid is used, 
this grid is only an intermediate step which does not impose periodic 
boundary conditions, and whose discreteness effects can be compensated 
using multigrid and direct summation methods
as described below.  Our approach is thus conceptually very
different from pixelisation approaches. It is also
useful at this stage to point out the obvious analogy between our
approach and gravity solvers of PM (particle - mesh) N-body simulations.

While the above approach works well for large scale modes, the
pixelisation introduces inaccuracies at scales comparable to the pixel
resolution. Our analogy with N-body simulations come to the rescue
here; PM simulations have similar inaccuracies on small scales that can
be corrected by introducing a direct summation between pairs of
particles at small seperations (P$^{3}$M - particle-particle
particle-mesh simulations). We start by splitting the convolution kernel
into short and long range pieces,
\be
\zeta \, = \, \zeta_{long} + \zeta_{short} \,= \,f(r)\zeta + g(r)\zeta
\,\,,
\ee
where $f$ and $g$ are filter functions with the properties that
\bea 
g = 1 \,\,\, (r < r_{min}) \nonumber \,\,\,,\\
f = 1 \,\,\, (r > r_{max}) \nonumber \,\,\,,\\
f = 0 \,\,\, (r < r_{min}) \nonumber \,\,\,,\\
g = 0 \,\,\, (r > r_{max}) \nonumber \,\,\,,\\
f + g = 1 \,\,\, (r_{min} < r < r_{max}) \,\,\, ,
\eea
with the definitions of long and short range are determined by $r_{min}$
and $r_{max}$. The multiplications are now done in two stages; the first
is to do the long range piece by the Fourier method described above,
while the short range correlations are done by direct summation.
Possible filter functions have the form $1-\cos^{2}(\theta)$ in the
regime $r_{min} < r < r_{max}$, where $\theta$ is an appropriately
scaled length. This form is chosen to minimise the inaccuracies that
result from the truncation of the correlation function. 

A second scheme to improve the resolution is to use a multi-grid approach.
The single-grid scheme described above is only the simplest such
implementation. One can trivially generalise it to multiple scales by
introducing a series of filters, $f_{i}$. In such implementations, the
direct summation is performed only for the shortest range; all other
convolutions are done by the Fourier method with the grid becoming
coarser for larger scales. Our codes used 3 scales - the innermost scale 
for direct summation, an intermediate scale with twice the resolution, and a 
coarse grid for the largest scales.

Note that while the $O(N \log N)$
scaling is dependent on a flat space assumption, it can be generalized 
to the sphere with no significant loss of efficiency. This is because
on a sphere only the coarse grid needs to be done with the slower 
$O(N_{\rm coarse}^{3/2})$ scaling, 
while the subsequent grids can still be 
$O(N_{\rm grid} \log N_{\rm grid})$ as long as the flat sky approximation holds on these 
grids. 

\subsection{Matrix Inversions}
The problem we focus on is of the form
\be
{\mathbf y} = C {\mathbf x} = (S+N) {\mathbf x} \,\,,
\ee
where ${\mathbf y}$ is known, $S$ is the signal matrix and $N$ is the
noise. We will assume that $N$ is diagonal in real space here, although
the algorithm can be modified for the case of sparse $N$. Also, we
assume that multiplication by $S$ can be efficiently performed by the
methods above. The matrix inversion is then performed by an underrelaxed
Jacobi iteration,
\be
{\mathbf x^{n+1}} = {\mathbf x^{n}} + (\overline{S} I + N)^{-1} \left[
{\mathbf y} - (S+N) {\mathbf x^{n}} \right] \,\,,
\ee
where $I$ is the identity matrix, and $\overline{S}$, the
underrelaxation parameter, is equal to $e_{max}$, the maximum eigenvalue
of $S$.  This can be estimated using the iterative scheme,
\bea
{\mathbf x^{n+1}} = \frac{S {\mathbf x^{n}}}{\parallel {\mathbf x^{n}}
\parallel} \nonumber \\
e_{max} = \parallel {\mathbf x^{n}} \parallel \,\, ,
\eea
which can intuitively be verified by considering the case of $S$
diagonal. 

The underrelaxed Jacobi iteration is only one of a number of possible approaches to computing the matrix inverses. For our matrices, we obtained a fractional precision of $10^{-4}$ in $\sim 100$ iterations and were therefore not limited 
by it. However, for ill-conditioned matrices, it might be necessary to go to conjugate gradient or multigrid methods.

\subsection{Trace estimation}
The only operation that cannot be trivially written in terms of
matrix-vector operations is the computation of the Fisher matrix (Eq. \ref{eq:fisher}) which
involves the trace of the product of four matrices. This is an intrinsically 
$O(N^3)$ process. 
If we assume a
random ensemble of vectors, ${\mathbf x}$, with the property that,
\be
\langle \mathbf{x x}^{t} \rangle = I,
\ee
we can use the following statistical
identity to estimate the trace,
\be
tr(A) = tr(AI) = tr(A\langle \mathbf{ x x}^{t} \rangle)  = {\mathbf x} A
{\mathbf x}^{t} \,\, .
\ee
The estimator is completely determined by specifying the ensemble
${\mathbf x}$. Note that if we choose an ensemble of $dim(\mathbf{x})$ orthogonal vectors, then we can exactly recover the trace; however, taking the  trace would then be an $O(N^{2}\log N)$ operation. This is already a gain from $O(N^3)$ and 
is achieved by the fast $O(N\log N)$ convolution methods of $C\mathbf{x}$ and $C^{-1}\mathbf{x}$ operations. However, 
$O(N^{2}\log N)$ is still slow compared to other operations. 
One might attempt to modify this by choosing a smaller subset of random, but orthogonal, vectors; we however find that this is slow to converge to the correct value. We obtain best convergence 
by using the real stochastic $Z_{2}$ esimator, where $\mathbf{x}$ is a random vector consisting of 1's and -1's. For $N = 90,000$, we measure the trace with a fractional precision of $\sim 10^{-5}$ with an ensemble of 400 random vectors.
This means that we have reduced the scaling from $O(N^3)$ to $O(N\log N)$ only.

\end{document}